\documentclass[twocolumn,nofootinbib]{revtex4}
\usepackage{amsmath,latexsym,revsymb,graphicx,epsfig,amssymb,verbatim,color}

\def\th{^{\mbox{\scriptsize th}}}
\def\zu{\{0,1\}}
\def\e{_\mathrm{e}}
\def\k{_\mathrm{r}}
\def\c{_\mathrm{c}}
\def\s{_\mathrm{s}}

\begin{document}

\title{Full security of quantum key distribution from no-signaling constraints}

\author{Llu\'{\i}s Masanes}
\affiliation{H.H.Wills Physics Laboratory, University of Bristol, Tyndall Avenue, Bristol BS8 1TL, U.K. email: ll.masanes@bristol.ac.uk}

\author{Renato Renner}
\affiliation{Institute for Theoretical Physics, ETH-Z\"urich, CH-8093, Z\"urich, Switzerland}

\author{Matthias Christandl}
\affiliation{Institute for Theoretical Physics, ETH-Z\"urich, CH-8093, Z\"urich, Switzerland}

\author{Andreas Winter}
\affiliation{
F\'{\i}sica Te\`orica: Informaci\'o i Fenomens Qu\`antics,
Universitat Aut\`onoma de Barcelona, ES-08193 Bellaterra (Barcelona), Spain. \\
ICREA-Instituci\'o Catalana de Recerca i Estudis Avancats,
Pg.~Lluis Companys 23, ES-08010 Barcelona, Spain.}

\author{Jonathan Barrett}
\affiliation{Department of Computer Science,
University of Oxford, Parks Road, Oxford OX1 3QD, U.K.\\}

\date{\today}

\begin{abstract} We analyze a cryptographic protocol for generating a distributed secret key from correlations that violate a Bell inequality by a sufficient amount, and prove its security against eavesdroppers, constrained only by the assumption that any information accessible to them must be compatible with the non-signaling principle. The claim holds with respect to the state-of-the-art security definition used in cryptography, known as universally-composable security. The non-signaling assumption only refers to the statistics of measurement outcomes depending on the choices of measurements; hence security is independent of the internal workings of the devices --- they do not even need to follow the laws of quantum theory. This is relevant for practice as a correct and complete modeling of realistic devices is generally impossible. The techniques developed are general and can be applied to other Bell inequality-based protocols. In particular, we provide a scheme for estimating Bell-inequality violations when the samples are not independent and identically distributed.  \end{abstract} \maketitle
	
\section{Introduction}

Quantum Key Distribution (QKD)~\cite{BB84,E91} is the task of generating a secret key such that the key gets known exclusively to two designated parties, in the following called \emph{Alice} and \emph{Bob}. In this work, we consider \emph{entanglement-based QKD}~\cite{E91}, where Alice and Bob have access to a source of entanglement, and they can communicate over an authenticated~\footnote{An authenticated channel provides the guarantee to the receiver, say Bob, that the received messages has indeed be sent by the sender, Alice. However, such a channel does not guarantee any secrecy.}  classical communication channel. Crucially, nothing is assumed about the source. In particular, the source may be fully controlled by an adversary, \emph{Eve}, who may try to gain information about the generated key.  The main idea behind entanglement-based QKD is that Alice and Bob, in a verification step, check whether the entanglement obtained from the source is sufficiently strong. If this is the case, they ``distill'' their key from the entanglement provided by the source. Otherwise, if the entanglement is found to be weak, they have to abort the protocol as security cannot be guaranteed.

The aim of this paper is to prove security of a class of QKD protocols under minimal assumptions. In particular, our argument is \emph{device-independent}~\cite{AGM,Pironio}, which means that we do not make any assumptions about the internal workings of the source nor the quantum devices used by Alice and Bob. In fact, we do not even require that they work according to the laws of quantum mechanics. Instead, we only make \emph{no-signaling assumptions}, which require that the components of the QKD scheme do not emit any undesired information~\cite{BHK}. This assumption can be met, for example, by perfectly isolating a large number of devices.  We also note that, in order to make sense of the QKD problem, certain non-signaling assumptions are necessary: If the device processing the final key broadcasts it, secrecy obviously remains unachievable in any protocol.

In contrast to this, most standard security proofs~\cite{LoChau, Mayers, ShorPreskill, Bihametal, ChReEk, ReGiKr03} are device-dependent. This means that the security claim is only guaranteed to hold if Alice and Bob's devices precisely meet a given specification. Consider, for example, an optical implementation, where a source distributes pairs of entangled photons, whose polarization is then measured by Alice and Bob. A device-dependent security proof would then require that the measurement outcomes depend in a specific way on the polarization degree of freedom, and are otherwise independent of any further properties of the photons (such as their wave length or their arrival time) or any other parameter (such as the temperature of the device). This assumption|even if one tolerates some inaccuracies|is not only hard to meet, but also hard to verify. In fact, the assumption is not met by many existing practical realizations of QKD, as has recently been demonstrated in a series of hacking experiments (see, e.g., \cite{Lo,Makarov}). 

The mismatch between the theoretical specifications of the devices used for the security proofs and the actual practical implementations of these devices was already recognized in the late 90s. In particular, Mayers and Yao proposed the idea of \emph{self-testing}, where the violation of Bell inequalities~\cite{Bell} is used to infer that the devices meet a given specification~\cite{MayersYao}.  This approach has been taken further by Barrett, Hardy, and Kent, who proposed a scheme whose security is based solely on certain non-signaling assumptions~\cite{BHK}, similarly to those used in this work. Their proof, however, only applied to an idealized setting with a noiseless source. Later, Ac\'{\i}n, Gisin and Masanes showed that noise can be tolerated if one makes the assumption of \emph{individual attacks}\footnote{An attack is called \emph{individual} if the adversary gathers only classical information, obtained by individual measurements applied to single pairs emitted by the source of entanglement.}~\cite{AGM}. We stress that these security claims, as well as the one presented in this work, do not rely on the correctness of quantum theory. 

In a parallel line of research, device-independent security proofs have been developed which rely on the validity of quantum theory. While the first proofs of this type were restricted to a certain class of attacks, called \emph{collective attacks}\footnote{An attack is called \emph{collective} if it is guaranteed that the individual particle pairs as received by Alice and Bob are independent and identically distributed~\cite{Bihametal}.}~\cite{ABGMPV}, this restriction could be relaxed recently to non-signaling constraints similar to those used here~\cite{Pironioetal,HanRen}. Moreover, these recent results, like the one presented here, use a strong notion of security, as introduced in~\cite{RK,BO,lluis}. This guarantees \emph{universal composability}, which means that the secret key generated by the protocol can safely be used in any application. %(see for a discussion).

All results mentioned above rely on the no-signaling assumption. In our model, we treat each measurement as if it was carried out on a separate device, between which no signaling is allowed.  In a practical setup---where Alice and Bob each have only one measurement device that is used repeatedly---this means that there should not be signaling between the individual uses of the devices. In particular, the non-signaling assumption is satisfied if the practical devices have no memory. This assumption has recently been relaxed in~\cite{PMAL, BaCoKe, VaVi}. Although the protocol in~\cite{PMAL} assumes that the adversary does not have a long term non-classical memory, \cite{BaCoKe} does not tolerate any noise, and the protocol in~\cite{VaVi} relies on the validity of quantum theory.

Since our claims are supposed to hold independently of the correctness of quantum theory, we need a general framework to specify the protocol, executed by Alice and Bob, as well as Eve's attack strategy. We follow a standard approach~\cite{prbox, ns, MAG} and describe the source of entanglement as a \emph{random system}, which takes inputs $X$, $Y$, and $Z$, and produces outputs $A$, $B$, and $E$, accessible to Alice, Bob, and Eve, respectively. The idea is that the inputs specify the choices of measurements that can be applied to the (potentially) correlated systems, and the outputs correspond to the measurement outcomes. Crucially, we assume that this random system satisfies certain non-signaling constraints.

The paper is organized as follows. In Section~\ref{sec:prel} we provide an introduction for: non-signaling correlations, Bell inequalities and their relation to privacy, the notion of device-independent protocols, and the security definition that we use. In Section~\ref{sec:protocol} we describe the protocol, state our main result, explain how to implement the protocol with quantum devices, and compare its performance with other protocols. The proofs of our results are provided in Section~\ref{sec:proof}. The conclusions of our work are in Section~\ref{conc}, and the Appendix contains a supplementary result that allows to adapt our protocol so that higher key rates can be achieved at the price of assuming the validity of quantum theory.

\section{Preliminaries} \label{sec:prel}

\subsection{Non-signaling correlations}

We use upper-case $A$ to denote the random variable whose particular outcome is the corresponding lower-case $a$. We write the probability distribution for $A$ as $P_A$, and the probability for $A=a$ as $P_a$, as well as $P_A (a)$. We use bold letters to denote strings of outcomes ${\bf a}= (a_1,\ldots, a_N)$ or strings of random variables ${\bf A}= (A_1,\ldots ,A_N)$.

Alice and Bob share $N$ pairs of physical systems, labeled by $n\in\{1,\ldots, N\}$. Alice measures her $n\th$ system with one of the $M$ observables $X_n \in \{0,1,\ldots,M-1\}$, obtaining the outcome $A_n \in \zu$. Analogously, Bob measures his $n\th$ system with one of the $(M+1)$ observables $Y_n \in \{0,1,\ldots, M\}$ and obtains the outcome $B_n \in \zu$. The chosen observables and their corresponding outcomes for the $N$ pairs of systems are represented by the random variables ${\bf A}$, ${\bf B}$, ${\bf X}$, ${\bf Y}$, which are correlated according to the joint conditional probability distribution $P_{{\bf A}, {\bf B}|{\bf X}, {\bf Y}}$. The number $P_{{\bf a} ,{\bf b}|{\bf x},{\bf y}}$ is the probability of obtaining the strings of outcomes ${\bf a}, {\bf b} \in \zu^{N}$ when measuring ${\bf x}\in \{0,\ldots, M-1\}^{N}$ and ${\bf y}\in \{0,\ldots,M\}^N$. The only assumption about this distribution is the following.

\medskip \noindent {\bf The non-signaling assumption: } {\em The choice of observable for one system cannot modify the marginal distribution for the rest of the systems.}

\medskip\noindent More formally, we impose the following condition among any two sets of input/output pairs ${\bf I}_1$, ${\bf O}_1$ and ${\bf I}_2, {\bf O}_2$,
\begin{equation}\label{ns1}
	\sum_{{\bf o}_2}P_{{\bf o}_1, {\bf o}_2| {\bf i}_1, {\bf i}_2}=
	\sum_{{\bf o}_2} P_{{\bf o}_1, {\bf o}_2| {\bf i}_1, {\bf i}_2'}
\end{equation}
for all ${\bf i}_2, {\bf i}_2', {\bf o}_1, {\bf i}_1$. 
%Although the two sets of subsystems are arbitrary, the above constrains turn out to be equivalent to the ones where $(O_2,I_2)$ corresponds to a single subsystem $(A_n,X_n)$ or $(B_n,Y_n)$. 
This condition could be enforced physically by using a different device for each measurement and isolating the devices from each other. In a more practical situation where the same device is used for subsequent measurements, the condition holds, for instance, if one assumes that the device has no memory. 

The information available to the adversary Eve is modeled analogously: it is given by the input/output behavior of a system, which may be correlated to Alice and Bob's measurements. Specifically, Eve can choose an observable $Z$ and obtains an outcome $E$.  We assume that this, together with the public messages exchanged by Alice and Bob, is all information available to her.  Note that we can without loss of generality assume that Eve only carries out this one measurement, for the sole assumption that we use in the security proof is that the global $(2N+1)$-partite distribution $P_{{\bf A}, {\bf B}, E|{\bf X}, {\bf Y}, Z}$ is a non-signaling one, but otherwise may be arbitrary.

%\medskip It is important to stress that systems inside Alice's laboratory must not signal each other, and the same for Bob. This may be quite difficult to implement in practice, but it is, in principle, possible.

\subsection{Bell inequalities}

\medskip A bipartite conditional distribution $P_{A,B|X,Y}$ is said to be local if it can be written as
\begin{equation}\label{local co}
  P_{a,b|x,y}^\mathrm{local} = \sum_v P_v\, P_{a|x,v} P_{b|y,v}\ , 
\end{equation}
for some probability distribution $P_V$ and conditional probability distributions $P_{A|X,V}$ and $P_{B|Y,V}$.
Local distributions can be generated by shared randomness (denoted $V$ above) between the parties, plus local operations. In other words, local distributions can be generated with classical resources. A distribution $P_{A,B|X,Y}$ which cannot be written as (\ref{local co}) is said to be non-local. Non-local correlations are the resource consumed in our secret key distribution protocol.

By definition, Bell inequalities \cite{Bell,CHSH,BC} are satisfied by all local distributions (\ref{local co}). In this paper, we concentrate on the Braunstein-Caves Bell inequality~\cite{BC}, or BC-inequality for short. This inequality is often stated using a different notation (not to be further used in this work), where $A_x, B_y$ denote the random variables $A,B$ conditioned on $X=x,Y=y$, so that it reads \begin{eqnarray}\nonumber
  &&\langle A_1 \oplus B_1 \rangle +
  \langle B_1 \oplus A_2 \rangle +
  \langle A_2 \oplus B_2 \rangle + \cdots 
\\ \label{BC original} &+&
  \langle A_M \oplus B_M \rangle +
  \langle B_M \oplus A_1 \oplus 1 \rangle
  \geq 1\ ,
\end{eqnarray}
where $\oplus$ is the sum modulo $2$.
For our purposes it is convenient to write the BC-inequality for a given conditional distribution $P_{A,B|X,Y}$ as the expectation of the random variable
\begin{equation}\label{var B}
  W = (A\oplus B \oplus \delta_X^0 \delta_Y^{M-1})
\end{equation}
over $P_{a,b,x,y} = P_{a,b|x,y} Q_{x,y}$, where 
\begin{equation}\label{Q}
  Q_{x,y} = \left\{ \begin{array}{cc}
    \frac{1}{2M} & \mbox{ if } (x-y \bmod M) \in \zu \\
    0 & \mbox{ otherwise}     
  \end{array} \right. .
\end{equation}
The BC-inequality for $M$ observables~\eqref{BC original} can be written as
\begin{equation}\label{BC inequality}
  \langle W\rangle \geq \frac{1}{2M} \ .
\end{equation}
As mentioned above, any local distribution (\ref{local co}) satisfies (\ref{BC inequality}). The largest violation $\langle W\rangle =0$ can be reached for certain non-signaling distributions, but cannot be reached within quantum theory. The largest quantum violation is obtained with the EPR state $|\phi\rangle = (|0\rangle |0\rangle + |1\rangle |1\rangle)/\sqrt{2}$ \cite{EPR, wehner}, with the measurements specified in FIG.~\ref{cercle}, reaching the value 
\begin{equation}\label{singlet}
  \langle W \rangle_{|\phi\rangle} = \sin^2 \left( \frac{\pi}{4 M} \right) \ .
\end{equation}
Note that when increasing $M$, the quantum violation tends to zero, the maximal one.

For $M=2$ the BC-inequality is equivalent to the famous CHSH-inequality~\cite{CHSH},
%\begin{equation} \langle A\oplus B \oplus \delta_{X=M-1} \delta_{Y=0} \rangle \geq \frac 1 4\ , \end{equation}
with its well-known maximal quantum violation of $  \langle W\rangle = \frac 1 2 - \frac{1}{2\sqrt{2}}\approx 0.15$ due to Tsirelson~\cite{cirelson}.

\subsection{Guessing probability and Bell violation}

Suppose that Eve is correlated with Alice and Bob through the global non-signaling distribution $P_{A,B,E|X,Y,Z}$. If Alice measures $X=0$ and obtains the outcome $A$, then we can quantify the knowledge that Eve has about $A$ by the optimal guessing probability
\begin{equation}
  \mathcal{P}_\mathrm{guess} (A|E) = \max_z \sum_e \max_a P_{A,E|X,Z} (a,e,0,z)\ .
\end{equation}
If $\mathcal{P}_\mathrm{guess} (A|E)=1$ then Eve knows $A$ with certainty. If $\mathcal{P}_\mathrm{guess} (A|E)=1/2$ then Eve is completely ignorant about the value of $A$. In \cite{BKP} it was shown that the knowledge that Eve has about $A$ can be bounded by the amount of non-locality present in Alice's and Bob's marginal distribution: 
\begin{equation}\label{aaa}
  \mathcal{P}_\mathrm{guess} (A|E) \leq 
  1/2 + M \langle W \rangle \ .
\end{equation}
If the marginal for the honest parties $P_{A,B|X,Y}$ violates the BC-inequality (\ref{BC inequality}), then according to (\ref{aaa}), the probability that Eve guesses correctly is smaller than one. 
%This is the reason why the Bell inequality (\ref{BC inequality}) is unconventionally written as a lower bound: the more non-locality the honest parties share, the lower $\langle \b \rangle$ is, and the less knowledge Eve has (\ref{aaa}). 
This is one manifestation of the monogamy of non-local correlations \cite{MAG,ns}. In Appendix~\ref{mnc}, inequality (\ref{aaa}) is generalized to the case of more than one pair of systems.

\subsection{Device-independent QKD}

Inequality~\eqref{aaa} allows to bound Eve's knowledge about $A$ in terms of the statistics of $A,B,X,Y$, regardless of how the correlations $P_{A,B|X,Y}$ are generated. In particular, the privacy of $A$ is independent of the functioning of the device used to generate $A$. Even if the devices are maliciously designed by Eve, and even if the devices violate quantum theory, the security of our protocol is not compromised. 

The only assumption that we make on the devices is that they satisfy the no-signaling constraints~\eqref{ns1}. This could be enforced by performing each of the $2N$ measurements by Alice and Bob in a separate isolated device.  Clearly, this approach, though theoretically possible, would be extremely costly in practice. A cheaper possibility|actually the one employed by all existing experiments|is that Alice and Bob each use one single device repeatedly for the different measurements. The constraints~\eqref{ns1} then mean that there should be no signaling between the individual uses of the devices. This would be the case, for instance, if the devices had no memory. While such a no-memory assumption may be hard to guarantee in practice, it is still considerably weaker than the assumption that the devices can be modeled completely, which is necessary in standard (non device-independent) cryptography.

\subsection{Security definition}\label{secdef}

Our key generation protocol starts from correlations that violate the Bell inequality~\eqref{BC inequality}. We model these initial correlations by a non-signaling distribution $P_{{\bf A,B},E|{\bf X,Y},Z}$. This distribution is a priori unknown and may have been chosen by the adversary. In other words, all our security claims are supposed to hold for any possible initial non-signaling distribution $P_{{\bf A,B},E|{\bf X,Y},Z}$.  Furthermore, Alice and Bob have access to a public authenticated communication channel. That is, all information sent through this channel will be available to, but cannot be altered by Eve.

The key distribution protocol specifies by a sequence of instructions for Alice and Bob. In each of the protocol steps, Alice and Bob either access the correlated data, perform local calculations, or exchange messages over the public channel. In the final step of the protocol, Alice and Bob generate the keys $K_A$ and $K_B$ taking values on $\{0,1\}^{N\s}$, respectively.
%More precisely, $K_A$ and $K_B$ are either bit strings of length $N\s$, or a special symbol, $\perp$, indicating that no key has been generated (this is for example the case if Alice and Bob find that the initial correlations are not sufficiently non-local). 

We say that a protocol is secure if the resulting distribution is indistinguishable from an ideal one. Suppose that at the end of the protocol all the relevant information is characterized by a distribution $P_{K_A K_B,T,E|Z}^\mathrm{real}$, where $T$ is a transcript of the communication, containing all messages exchanged between Alice and Bob through the authenticated channel (note that $T$ is accessible to Eve). An ideal QKD protocol produces the distribution \begin{equation}\label{ideal}
  P_{k_A k_B,t,e|z}^\mathrm{ideal} 
\ =\  
  2^{-N\s}\, \delta_{k_A}^{k_B}\,
  P_{t,e|z}^\mathrm{real}\ ,
\end{equation}
where $P_{t,e|z}^\mathrm{real}$ are the values of the marginal $P_{TE|Z}^\mathrm{real}$ derived from the real distribution $P_{K_A K_B T E|Z}^\mathrm{real}$. Note that according to~\eqref{ideal}, the two versions of the secret key, $K_A$ and $K_B$, are identical and uniformly distributed, independently of the values taken by $T,E,Z$. We say that a protocol is secure if the quantity
\begin{equation} \label{eq_secretkey}
  \sum_{k_A, k_B, t} \max_z \sum_e 
  \Big| P_{k_A k_B,t,e|z}^\mathrm{real} -
  P_{k_A k_B,t,e|z}^\mathrm{ideal} \Big|
\end{equation}
can be made arbitrarily small as $N$ grows. This is the strongest notion of security, and it is called {\em universally composable security}~\cite{RK, BO, Renner, QuadeRen}. It is often the case that the secret key generated by a QKD protocol is used as an ingredient for another cryptographic task. The above security definition warrantees that the composed scheme that uses a secure key distribution protocol as a component is as secure as if an ideal secret key~\eqref{ideal} was used instead (see~\cite{lluis} for more details).

\section{Setup and results} \label{sec:protocol}

\subsection{Description of the protocol}\label{protoc}

In what follows we describe a family of protocols parametrized by the number of settings $M$ in the BC-inequality, and a probability $\gamma \in (0,1)$. The role of $\gamma$ is explained next. In Section~\ref{IIIB} we illustrate how to find the optimal value for these parameters. This family of protocols is similar to those introduced in~\cite{BHK,AMP}.

\medskip\noindent {\bf 1. Distribution and measurements.} Alice and Bob are given $N$ pairs of systems. Alice generates the random bits ${\bf I}= (I_1,\ldots, I_N)$ independently and with identical distribution: $P_{I_n} (0) =1-\gamma, P_{I_n}(1) =\gamma$. Analogously, Bob generates the random bits ${\bf J}= (J_1,\ldots, J_N)$ independently and with identical distribution $P_{J_n} =P_{I_n}$. Pairs such that $I_n =J_n =0$ are used to generate the raw key, and pairs such that $I_n =J_n =1$ are used to estimate how much non-locality Alice and Bob share. For each $n\in\{1,\ldots, N\}$, if $I_n=0$ Alice measures her $n\th$ system with $X_n=0$, if $I_n=1$ she measures it with $X_n$ chosen uniformly on $\{0,\ldots, M-1\}$, if $J_n=0$ Bob measures his $n\th$ system with $Y_n=M$, if $J_n=1$ he measures it with $Y_n$ chosen uniformly on $\{0,\ldots, M-1\}$.

\medskip \noindent {\bf 2. Estimation of non-locality.} Alice and Bob announce ${\bf I}, {\bf J}$ publicly as well as the tuples $(A_n, B_n, X_n, Y_n)$ for the values of $n$ where $I_n=J_n=1$. The subset of pairs \begin{align}\nonumber
	{\cal N}\e = &\Big\{n\in\{1,\ldots , N\}\ |\ I_n=J_n=1 
\\ \label{condition}
	&\mbox{ and } (X_n -Y_n \bmod M) \in\{0,1\} \Big\}\ .
\end{align}
is used to compute the average value for the BC-inequality
	\begin{equation}\label{est non-locality}
   \bar W = \frac{1}{N\e} 
	\sum_{n\in {\cal N}\e} (A_n\oplus B_n \oplus \delta_{X_n}^{0} \delta_{Y_n}^{M-1}) \ ,
	\end{equation}
        where $N\e= |{\cal N}\e|$. $\bar W$ thus corresponds to the average of the variables $W_n$ defined by~(\ref{var B}) with $n\in {\cal N}\e$. Note that after post-selecting on the pairs with $n\in {\cal N}\e$ the random variables $X_n, Y_n$ follow the distribution $Q_{X,Y}$ defined by~(\ref{Q}), which allows to identify $W_n$ with the BC-inequality for the pair with index~$n$.

The number of estimated systems is $N\e \approx 2 N \gamma^2/M$ with high probability. Here and in the rest of the paper the symbol $\approx$ denotes equality up to subleading terms. As we will see, the asymptotic efficiency of the protocol does not depend on the subleading terms. The outcomes of the pairs in the set
\begin{equation}\label{raw set}
	{\cal N}\k = \{n \in\{1,\ldots , N\}\ |\ I_n=J_n=0 \}\ ,
\end{equation}
have not been published, and are denoted by ${\bf A}\k, {\bf B}\k$. These are the raw keys obtained by Alice and Bob, respectively. We denote their length by $N\k = |{\cal N}\k| \approx (1-\gamma)^2 N$.

\medskip\noindent {\bf 3. Error correction.} Alice publishes $N\c$ bits of information about her raw key $C= f({\bf A}\k)$, which Bob uses for correcting the errors in his raw key: $({\bf B}\k, C) \mapsto {\bf B}\k' \approx {\bf A}\k$. Any error-correction method, or equivalently any function $f:\zu^{N\k} \to \zu^{N\c}$, can be inserted here, as long as the probability that $\mathbf{B}\k' \neq {\bf A}\k$ vanishes as $N$ grows. It follows from classical information theory (see~\cite{Renner} for more details) that error correction can be achieved asymptotically with \begin{equation}\label{eeee}
  N\c \approx N\k h(\lambda)\ ,
\end{equation}
where $\lambda$ is the relative frequency of $B_n \neq A_n$ for all $n\in {\cal N}\k$, and
\begin{equation}
  h(\lambda) = -\lambda \log_2 \lambda -(1-\lambda) \log_2 (1-\lambda)
\end{equation}
is the binary Shannon entropy.

\medskip\noindent {\bf 4. Privacy amplification.} Alice chooses at random a function $G: \zu^{N\k} \rightarrow \zu^{N\s}$ from a set of two-universal hash functions (see Definition \ref{2universal} or \cite{BBCM94}) with output length \begin{eqnarray}\nonumber
  N\s (\bar w) &=& \max\Big\{0,
\\ \nonumber &&  
  \max_{\theta \in [0,1]} \big[ 2 N\e D(\bar w\|\theta) -2 N\k \log_2 (1/2 + M\theta)\big]
\\ \label{Ns 12} &&
   -N\k -N\c  -2\log_2 (8 N\e N /\epsilon) 
   \Big\} \ ,
\end{eqnarray}
where the binary relative entropy is defined as
\begin{equation}\label{relent}
  D(\theta_1 \| \theta_2) = \theta_1 \log_2 \frac{\theta_1}{\theta_2} + (1-\theta_1) \log_2 \frac{1-\theta_1}{1-\theta_2} \ ,
\end{equation}
and $\bar w$ is the observed value of the random variable~\eqref{est non-locality}.
If $N\s(\bar w) >0$ then Alice and Bob respectively compute $K_A= G(\mathbf{A}\k)$ and $K_B= G(\mathbf{B}\k')$, which constitute their versions of the final secret key. We stress that the hash function $G$ is chosen at random and independently of any other information.  
The first maximization in~\eqref{Ns 12} avoids a negative length for the secret key, which obviously does not have any meaning. If $N\s(\bar w) =0$ then Alice and Bob write $K_A=K_B= \perp$, which means that the protocol has not produced any secret key. 

\subsection{Main Results}

The above protocol can be seen as a 
process which transforms
process which transforms the initial distribution $P_{{\bf A,B},E|{\bf X,Y}, Z}$ into the final distribution $P_{K_A, K_B, \bar W, T, E|Z}$, where
\begin{equation}\label{T}
  T= \Big[ {\bf I,J}, C,G, (A_n, B_n, X_n, Y_n)\ \forall n\in {\cal N}\e \Big]
\end{equation}
is all the information that has been published by the honest parties.
We prove that for any initial distribution $P_{{\bf A,B},E|{\bf X,Y}, Z}$, the resulting distribution $P_{K_A, K_B, \bar W, T, E|Z}$ satisfies
\begin{eqnarray}
\nonumber &&
  \hspace{-3mm} \sum_{k_A, k_B, \bar w,t} 
  \hspace{-2mm} \max_z \sum_e 
  \Big| P_{k_A, k_B, \bar w,t,e|z} -2^{- N\s (\bar w)}\,  
  \delta_{k_A}^{k_B}\, 
  P_{\bar w,t,e|z} \Big|
\\ \label{18} &\leq &
	\epsilon + 2\epsilon_\mathrm{erco}\ ,
\end{eqnarray}
where $\epsilon_\mathrm{erco}$ is an upper-bound for the error probability of the error correction scheme. This implies that the actual key generated by the protocol has at most distance $\epsilon + 2\epsilon_\mathrm{erco}$ from a perfectly secure key (see also~(\ref{eq_secretkey}) above).  Note that the parameter $\epsilon$ can be controlled by the honest parties when adjusting the length of the final secret key~\eqref{Ns 12}.

By setting $\epsilon$ and $\epsilon_\mathrm{erco}$ to sufficiently small values, the honest parties can be confident of the fact that the secret key generated by the protocol is indistinguishable from an ideal secret key~\eqref{ideal}. This implies that the protocol is secure according to the strongest notion of security, the so called universally-composable security~\cite{QuadeRen,lluis} (see Section~\ref{secdef}).

%\begin{eqnarray}&& \nonumber  \sum_{k_A,k_B, \bar w,c,g} \hspace{-2mm}  \max_z \sum_e \Big|   P_{k_A, k_B,\bar w,c,e,g|z} -2^{-2 N\s (\bar w)}   P_{\bar w,c,e,g|z} \Big|\\ && \leq   \epsilon +\epsilon_\mathrm{erco}\ . \end{eqnarray}
The efficiency of a key distribution scheme is quantified by the asymptotic secret key rate. This is defined as the ratio $N\s/N$ in the limit $N\rightarrow \infty$, where $N\s$ is the number of perfect secret bits obtained and $N$ is the number of pairs of systems consumed. Using~\eqref{Ns 12} and~\eqref{eeee} we obtain the secret key rate of our protocol:
\begin{eqnarray}
\label{Ns 13}
  \lim_{N\to \infty} \frac{N\s}{N} = -(1-\gamma)^2[1+ h(\lambda)] +
\hspace{30mm}
\\ \nonumber +
  \max_{\theta \in [0,1]} \left[ \frac{4 \gamma^2}{M} D(\bar w\|\theta) -2 (1-\gamma)^2 \log_2\! \Big(\frac 1 2 + M\theta \Big)\right]
\end{eqnarray}
It is understood that if the above quantity is negative the secret key rate is zero.

\subsection{Implementation of the protocol with quantum devices}\label{IIIB}

Here we explain how to implement the protocol with quantum-mechanical devices, i.e., if the initial correlations are generated by measurements on an entangled quantum state. We stress that this is not a part of the proof that the resulting key is secret (as the secrecy claim must hold for any possible correlations).  However, it is necessary to argue that, if the adversary is passive then the protocol actually generates a key (in particular, it should not abort), and to calculate the rate it which it does so|thereby allowing a comparison to other protocols. 

\begin{figure}
\begin{center}
  \includegraphics[width=65mm]{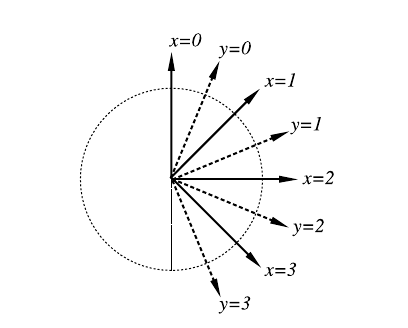}
  \caption{Location in the equator of the Bloch sphere of the observables for $M=4$.}
  \label{cercle}
\end{center} 
\end{figure}

\medskip Suppose Alice and Bob share many copies of the noisy EPR state
\begin{equation}\label{noise}
  \rho = (1-\xi)\, \Phi + \xi \frac{\mathbb{I}}{4}\ ,
\end{equation}
where $\xi\in [0,1]$ is the fraction of noise, $\Phi$ is the projector onto the EPR state $|\phi\rangle = (|0\rangle |0\rangle +|1\rangle |1\rangle)/\sqrt{2} $, and $\mathbb{I}/4$ is the maximally noisy state. They perform the measurements in the following orthogonal basis. The observable $x \in \{0,\ldots, M-1 \}$ for Alice has eigenvectors 
\begin{equation}\label{A basis}
  |0\rangle \pm e^{i\pi x/M} |1\rangle\ ,
\end{equation}
the observable $y \in \{0,\ldots, M-1 \}$ for Bob has eigenvectors
\begin{equation}\label{B basis}
  |0\rangle \pm e^{i\pi (y + 1/2)/M}|1\rangle\ ,
\end{equation}
and the observable $y=M$ for Bob has eigenvectors
\begin{equation}\label{rk}
  |0\rangle \pm |1\rangle\ .
\end{equation}
Note that, while Alice has $M$ observables, Bob has $M+1$ observables, and that $y=M$ is the same observable as Alice's $x=0$. In the Bloch sphere, these observables correspond to the directions represented in FIG.~\ref{cercle}. 
The observables $x,y\in \{0,\ldots, M-1\}$ are the ones used to obtain large violations of the BC-inequality \cite{BC}. The observables $x=0, y=M$ maximize the correlation between Alice and Bob, and hence, are used to generate the raw key. For $M=2$, this protocol is essentially equivalent to Ekert's original protocol~\cite{E91}.

\begin{figure}
\begin{center}
  \includegraphics[width=80mm]{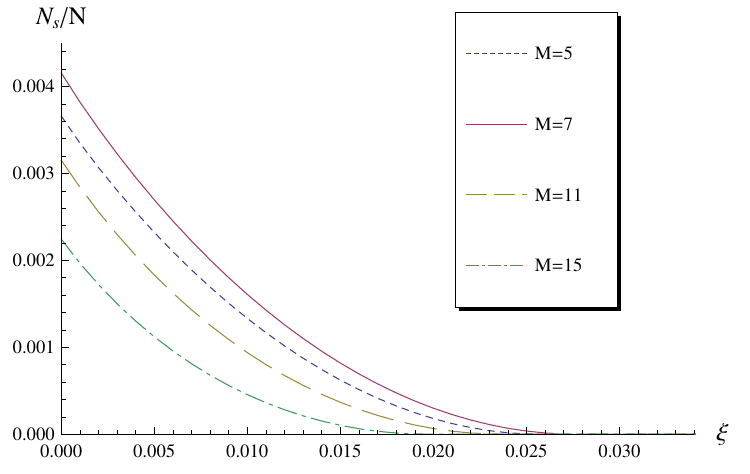}
  \caption{The secret key rate $N\s /N$ as a function of the noise $\xi$ is plotted for different values of $M$. The parameter $\gamma$ is optimized for every value of $\xi$. Note that in our protocol, $\gamma$ tends to a non-zero value asymptotically.
}
  \label{chshfig}
\end{center} 
\end{figure}

Using Equation~\eqref{singlet} it is straightforward to extend the expectation of the Bell inequality violation~\eqref{BC inequality} to the case where noise is added to the EPR state~\eqref{noise}:
\begin{equation}\label{W rho}
  \langle W \rangle_{\rho} = (1-\xi)\sin^2\! \left( \frac{\pi}{4 M} \right) + \xi\, \frac 1 2 \ .
\end{equation}
This is the value taken by $\bar W$ in the large $N$ limit. Substituting this in~\eqref{Ns 12} we obtain the secret key rate, which is plotted as a function of the noise $\xi$ in Fig.~\ref{chshfig}. The value of the parameter $\gamma$ is numerically optimized, such that, for each value of the noise $\xi$ the secret key rate is maximal.

In Fig.~\ref{chshfig} one can see that the secret key rate for $M=2$ (and also $M=3$) is zero, hence the security of the CHSH-protocol~\cite{AGM} against general attacks remains an open question. We have numerically seen that the optimal protocol is the one with $M=7$.

%For $M=3$ the rate is non-zero at high $p$, but quite small. For $M=6$ the protocol tolerates the maximum level of noise ($p_\mathrm{min} = .972$). Each amount of noise $p$ has an optimal number of observables $M$ which maximizes the rate. In the noiseless limit $p \rightarrow 1$ the optimal $M$ tends to infinite $M\rightarrow \infty$.

\subsection{Comparison with other results}

In this section we make a survey of some proofs for device-independent security in QKD. The first division that we consider is whether the adversary is totally unrestricted~\cite{BHK, Pironioetal, HanRen, PMAL, BaCoKe, VaVi}, or is constrained to perform the so called individual or collective attacks~\cite{AMP, AGM, ABGMPV, Pironio}. The second division concerns whether the security relies on the validity of quantum theory~\cite{ABGMPV, Pironio, Pironioetal, HanRen, VaVi} or not~\cite{BHK, AMP, AGM, PMAL, BaCoKe}. Third division, the devices are assumed to be memoryless~\cite{BHK, AMP, AGM, ABGMPV, Pironioetal, HanRen}, or not~\cite{PMAL, BaCoKe, VaVi}. Fourth division, the protocol tolerates certain degree of noise~\cite{AMP, ABGMPV, Pironio, Pironioetal, VaVi, AGM, HanRen, PMAL}, or not~\cite{BHK, BaCoKe}. Fifth division, the adversary is allowed to have a long-term non-classical memory~\cite{AMP, ABGMPV, Pironio, Pironioetal, VaVi, AGM, HanRen, BHK, BaCoKe}, or not~\cite{PMAL}.

The advantages of our result are: 
%\RR{As you were introducing the different ``divisions'' above, it would be natural to say more explicitly in which of the divisions our protocol and analysis is situated, rather than give this list of advantages / disadvantages. Otherwise, the two parts look a bit unconnected. But I assume there is some reason for this choice?}
\begin{enumerate}
\item The adversary is totally unrestricted in the sense that no assumption is made about the structure of the global distribution (like in individual or collective attacks).

\item The security of our protocol does not rely on the validity of quantum theory.

\item The adversary is allowed to have a long-term non-classical (and non-quantum) memory.

%\item It is compatible with any error-correction scheme.
\end{enumerate}
The disadvantages of our result are:
\begin{enumerate}
\item The security of our protocol relies on the measuring devices not having an internal memory.

\item Our protocol tolerates certain degree of noise, but only a small amount $\xi = 2.7\%$ (compare this with $\xi= 11\%$ in~\cite{PMAL}).

\item The secret key rate achieved by our protocol is very low. For example, in the noiseless case $\xi=0$, our protocol gives $N\s/N = .004$ while the protocols of~\cite{ABGMPV, Pironio, Pironioetal, AGM, HanRen, PMAL} obtain $N\s/N = 1$.
\end{enumerate}

%The only existing result which enjoys similar advantages is~\cite{HanRen}, which has the additional advantage of having a larger secret key rate. But the protocol in~\cite{HanRen} is not compatible with arbitrary error-correction schemes, and ours is.

\section{Security Proof} \label{sec:proof}

\newtheorem{defin}{Definition}
\newtheorem{lemma}[defin]{Lemma}
\newtheorem{theorem}[defin]{Theorem}
\newtheorem{corollary}[defin]{Corollary}

\subsection{Properties of symmetric distributions}
The results derived in this section provide tools for estimating properties of symmetric distributions without resorting to any de Finetti-like theorem. They are motivated by recent quantum analogues~\cite{post-selection} (sometimes termed \emph{postselection techniques}~\cite{post-selectionpedagogical}) and may be of independent interest.

We use calligraphic letter ${\cal V}$ to denote the alphabet of values for the corresponding random variable $V$, that is $v\in {\cal V}$. We use bold letters to denote strings of variables ${\bf v} = (v_1, \ldots , v_N) \in {\cal V}^N$ or random variables ${\bf V}= (V_1, \ldots , V_N)$. We say that a distribution $P_{\bf V}$ is symmetric if $P_{\bf V} (v_1, \ldots , v_N) = P_{\bf V} (v_{\pi (1)}, \ldots , v_{\pi (N)})$ for any permutation $\pi : \{1, \ldots ,N\} \to \{1, \ldots ,N\}$.
\begin{defin}\label{freq defi}
  Given a string ${\bf v} = (v_1,\ldots, v_N) \in \mathcal{V}^{ N}$ we define its corresponding frequency distribution $q= \mathrm{freq}({\bf v})$ as
  \begin{equation}\label{freq}
    q(v) = \frac{| \{n: v_n=v\}|}{N}\ ,
    \quad \forall\, v\in {\cal V}\ .
  \end{equation}
  This function naturally extends to sets ${\cal Q} = \mathrm{freq} ({\cal V}^{ N})$, and random variables $Q = \mathrm{freq} ({\bf V})$.
\end{defin}
For any ${\bf v}$, the frequency $q= \mathrm{freq}({\bf v})$ is a probability distribution for the random variable $V$, but it has the specific feature that it only takes values on the set \mbox{$\{ \frac{n}{N} : n=0,\ldots, N \}$}. $\mathcal{Q}$ is the set of all possible frequencies, whose cardinality can be bounded as
\begin{equation}\label{size}
    |\mathcal{Q}| \leq (N+1)^{|{\cal V}|-1}\ .
\end{equation}
For what follows, it is convenient to define a particular kind of probability distributions for $\mathbf{V}$: {\em the distribution with well-defined frequency} $q\in {\cal Q}$, denoted $P_{{\bf V}|q}$, is the uniform distribution over all strings ${\bf v} \in {\cal V}^{ N}$ such that $\mbox{freq}({\bf v}) =q$. Another important kind of symmetric distributions are the {\em i.i.d.~distributions}, representing independent and identically-distributed random variables $V_1,\ldots, V_N$. Borrowing notation from quantum information theory we write $P_V^{\otimes N}$ for such a distribution. If $P_V(v) < 1$ for all $v$, then the i.i.d.~distribution $P_V^{\otimes N}$ does not have a well-defined frequency. However, any symmetric distribution $P^\mathrm{sym} _{{\bf V}}$, including the i.i.d. ones, can be written as a mixture of distributions with well-defined frequency,
\begin{align}\label{ccc}
  P^\mathrm{sym} _{{\bf V}} &= \sum_{q\in\mathcal{Q}} P_Q(q)\, P_{{\bf V}|q}\ , \\
  Q &= \mathrm{freq}({\bf V})\ .
\end{align}
These two equalities establish a one-to-one correspondence between $Q$ and ${\bf V}$, for symmetric distributions. In Lemma~\ref{prop 1} we show that, in a sense, general symmetric distributions are similar to i.i.d.~distributions. But before, we need the following technical result.
%%%%%%%%%%%%%%
\begin{lemma}\label{renato}
    Let the probability distribution $P_V$ take values on the set $\{\frac{n}{N} : n=0,\ldots, N \}$, and let $\mathbf{V} = (V_1, \ldots, V_N)$ be distributed according to $P_V^{\otimes N}$.  Then the probability distribution $P_Q$ for $Q= \mathrm{freq} (\mathbf{V})$ takes its maximum at $Q= P_V$, that is,
    \begin{equation}
        P_Q(P_V) = \max_{q \in \mathcal{Q}} P_Q(q) \ .
    \end{equation}
\end{lemma}
{\em Proof }
We show that for any $q \in \mathcal{Q}$ with $q \neq P_V$ there exists $q' \in \mathcal{Q}$ such that $P_Q(q') > P_Q(q)$. Thus let $q \in \mathcal{Q}$ be fixed such that $q \neq P_V$. We call the \emph{support} of $q$: the set of values $v$ such that $q(v) > 0$. If the support of $q$ is not contained in the support of $P_V$ then $P_Q(q) = 0$.  We can thus without loss of generality assume that the alphabet of $V$, denoted $\mathcal{V}$, is the support of $P_V$, that is, $P_V (v) > 0$ for all $v \in \mathcal{V}$. For any $v \in \mathcal{V}$ define
\begin{equation*}
  d(v) = q(v) - P_V (v)\ .
\end{equation*}
Furthermore, let $v_{\min}$ and $v_{\max}$ be defined by
\begin{equation*}
  \begin{array}{ll}
  d(v_{\min}) = \min_v d(v) \\
  d(v_{\max}) = \max_v d(v)
  \end{array}\ .
\end{equation*}
Because $q\neq P_V$ and the assumption of the lemma, $d(v_{\min}) \leq - 1/N$ and \mbox{$d(v_{\max}) \geq 1/N$}. Let us define $q' \in \mathcal{Q}$ as
\begin{equation*}
  q'(v) = \begin{cases} q(v) + \frac{1}{N} & \text{if $v = v_{\min}$} \\  q(v) - \frac{1}{N} & \text{if $v = v_{\max}$} \\  q(v) & \text{otherwise.} \end{cases}
\end{equation*}
From  the two inequalities above we have
\begin{equation}
  \begin{array}{ll}\label{eq:qPminrel}
    q'(v_{\min}) \leq P_V (v_{\min})  \\
    q'(v_{\max}) \geq P_V (v_{\max})
  \end{array}\ .
\end{equation}
Using the identity
\begin{equation*}
  P_Q(q) =
  \frac{N! \prod_v P_V (v)^{q(v) N} }{\prod_v (q(v) N)!}
\end{equation*}
we find
\[
  \frac{P_Q(q')}{P_Q(q)}
=
  \frac{P_V (v_{\min}) (q'(v_{\max}) + \frac{1}{N})}{P_V (v_{\max}) q'(v_{\min})}
>
  \frac{P_V (v_{\min})}{P_V (v_{\max})} \frac{q'(v_{\max})}{q'(v_{\min})}
\]
(note that the terms in the denominator cannot be zero).
By~\eqref{eq:qPminrel}, the right-hand side cannot be smaller than $1$, which concludes the proof.
\hfill $\Box$

\begin{lemma}\label{prop 1}
    If there is a function $t: {\cal V}^{ N} \to \mathbb{R}$ and $\epsilon >0$ such that for any (single-copy) distribution $P_V$ the bound 
\begin{equation}\label{prem1}
  \sum_{\bf v} P_{V}^{\otimes N} ({\bf v})\, t({\bf v}) \leq \epsilon
\end{equation}
holds, then for any symmetric distribution $P^\mathrm{sym} _{{\bf V}}$ we have
\begin{equation}\label{cond S}
  \sum_{\bf v} P^\mathrm{sym}_{{\bf V}} ({\bf v})\, 
  t({\bf v}) \leq |{\cal Q}|\, \epsilon\ .
\end{equation}
\end{lemma}
{\em Proof } Let us first prove (\ref{cond S}) for distributions with well-defined frequency $P^\mathrm{sym}_{{\bf V}} = P_{{\bf V}|q}$, for all $q\in \mathcal{Q}$. 
Since any $q \in \mathcal{Q}$ is a (single-copy) distribution for $V$, the premise~\eqref{prem1} applies to it:
\begin{equation}
  \sum_{\bf v} q^{\otimes N}\! ({\bf v})\, t({\bf v}) \leq \epsilon
\end{equation}
Using the decomposition (\ref{ccc}), we know that there is a random variable $Q'$ such that $\sum_{q' \in\mathcal{Q}} P_{Q'}(q')\, P_{{\bf V}|q'} = q^{\otimes N}$, and then
\begin{equation}\label{j}
    \sum_{\bf v} \sum_{q' \in\mathcal{Q}} P_{Q'}(q') P_{{\bf V}|q'}({\bf v}) \, t({\bf v}) \leq \epsilon\ .
\end{equation}
In Lemma~\ref{renato} it is shown that the distribution $P_{Q'}(q')$ reaches the maximum at $q'= q$, which implies $P_{Q'}(q) \geq 1/|\mathcal{Q}|$. Then
\begin{align}\nonumber
    & \sum_{\bf v} P_{{\bf V}|q}({\bf v})\, t({\bf v}) \\
    \leq\ & \label{cond F}
    |\mathcal{Q}|\,  P_{Q'} (q)\, \sum_{\bf v}  P_{{\bf V}|q} ({\bf v})\, t({\bf v})
    \ \leq\
    |\mathcal{Q}|\, \epsilon \ ,
\end{align}
where the last inequality follows from (\ref{j}). Finally, we prove (\ref{cond S}) by applying the bound (\ref{cond F}) to each term in (\ref{ccc}). \hfill $\Box$\medskip

An equivalent way to write the above result for the case ${\cal V}= \zu$, which will be useful later, is the following. For any symmetric distribution $P^\mathrm{sym} _{{\bf V}}$ and any function $t$ we have
\begin{equation}\label{cond S 2}
  \sum_{\bf v} 
  P^\mathrm{sym}_{{\bf V}} ({\bf v})\, 
  t({\bf v}) 
\ \leq\  
  (N+1)
  %|{\cal Q}|\, 
  \max_{P_V}  \sum_{\bf v} 
  P_V^{\otimes N}\! ({\bf v})\, 
  t({\bf v})
\end{equation}
where the maximization is over single-copy distributions for $V$.

%The following lemma, proven in [16], establishes an important property for i.i.d. distributions.

%\begin{lemma} Let $V_1,\ldots ,V_N$ be i.i.d. random variables distributed according to $P_V$. The probability distribution for the frequency $Q = \mathrm{freq} (V_1, \ldots, V_N)$ can be bounded as
%\begin{equation} P_Q(q) \leq 2^{ -N D(q \| P_V)}\ ,\end{equation}
%where for any pair of distributions $q_1, q_2$, the relative entropy is defined as
%\begin{equation} D(q_1 \| q_2) = \sum_{v\in {\cal V}} q_1(v) \log_2 \frac{q_1(v)}{q_2(v)}\ .
%\end{equation}
%\end{lemma}

\subsection{Properties of non-signaling distributions}

For the following presentation it is useful to introduce some additional notation.  We represent single-pair distributions $P_{A,B|X,Y}$ as vectors with components arranged in the following way \begin{eqnarray}\label{arrangement}
  P_{A,B|X,Y} =\hspace{60mm}
  \\ \nonumber
  \scriptsize
  \begin{array}{|cc|cc|cc|}
    \hline
    P(0,0|0,0)&P(0,1|0,0) &\ldots &\qquad &P(0,0|0,M\!-\!1)&\qquad \\
    P(1,0|0,0)&P(1,1|0,0) & & & & \\
    \hline
    \qquad\vdots &&\ddots&&\vdots& \\
    &&&&& \\
    \hline
    P(0,0|M\!-\!1,0)&&\hdots & & P(0,0|M\!-\!1,M\!-\!1)&\qquad \\
    & & & & & \\
    \hline
  \end{array}
  \normalsize
\end{eqnarray}
Define the following two vectors (which are not probability distributions)
\begin{eqnarray}
  \mu &=&\frac{1}{4M}
  \begin{array}{|cc|cc|cc|cc|}
    \hline
    1 & 1 &   &   & & & 1 & 1 \\
    1 & 1 &   &   & & & 1 & 1 \\
    \hline
    1 & 1 & 1 & 1 &  &  &  & \\
    1 & 1 & 1 & 1 &  &  &  & \\
    \hline
    \phantom{\ddots}  &  &\ddots  &  &\ddots  &  &\phantom{\ddots}  &  \\
      &  &  &  &  &        &  &   \\
    \hline
      &   &  &  & 1  & 1  & 1 & 1 \\
      &   &  &  & 1  & 1  & 1 & 1 \\   \hline
  \end{array}\ , \\
  \nu &=&\frac{1}{2}
  \begin{array}{|cc|cc|cc|cc|}
    \hline
     0 & 1 & & & & & 1 & 0 \\
    -1 & 0 & & & & & 0 &-1 \\
    \hline
     0 &-1 & 0 & 1 & &  &  & \\
     1 & 0 &-1 & 0 &  &  &  & \\
    \hline
      &  &\ddots  &  &\ddots  &  &  & \\
      &  &  &\phantom{-1}  &  &\phantom{-1} & &     \\
    \hline
     &  &  &  & 0&-1& 0 & 1 \\
     &  &  &  & 1& 0&-1 & 0 \\
   \hline
 \end{array}\ ,
\end{eqnarray}
where empty boxes have to be understood as having zeros
\begin{equation}
	\begin{array}{|cc|}
		\hline
		\phantom{0} & \phantom{0} \\
		& \\
		\hline
	\end{array}
	=
	\begin{array}{|cc|}
	\hline
	0 & 0 \\
	0 & 0 \\
	\hline
	\end{array}\ ,
\end{equation}
and ellipsis between two identical boxes have to be understood as an arbitrarily large sequence of identical boxes. From now on, the absolute value of a vector means component-wise absolute value. For example 
\[
  |\nu| =\frac{1}{2}
  \begin{array}{|cc|cc|cc|cc|}
    \hline
     0 & 1 & & & & & 1 & 0 \\
     1 & 0 & & & & & 0 & 1 \\
    \hline
     0 & 1 & 0 & 1 &  &  &  & \\
     1 & 0 & 1 & 0 &  &  &  & \\
    \hline
    \phantom{\ddots}  &  &\ddots  &  &\ddots  &  &\phantom{\ddots}  & \\
      &  &  &  &  &        &  &     \\
    \hline
     & &  &  & 0 & 1 & 0 & 1 \\
     & &  &  & 1 & 0 & 1 & 0 \\
   \hline
 \end{array}\ .
\]
Also, an inequality "$\preceq$" between two vectors means components-wise inequality "$\leq$". For example $\nu \preceq |\nu|$. Define the vectors
\begin{eqnarray}\label{beta a}
	\beta_a &=& \mu + (-1)^a \nu \ , 
\\ \label{beta}
	\beta &=& \mu + |\nu| \ .
\end{eqnarray}
One can check that the Braunstein-Caves Bell inequality, defined in~\eqref{var B} and~\eqref{BC inequality}, can be written as 
\begin{equation}\label{id BC}
	\beta\cdot P_{A,B|X,Y} = \frac 1 2 + M\langle W \rangle \geq  1 \ .
\end{equation}
Above, the symbol ``$\cdot$" represents the scalar product between the vectors $\beta$ and $P_{A,B|X,Y}$.

\begin{lemma}\label{lemma 1}
If $P_{{\bf A}, {\bf B}|{\bf X}, {\bf Y}}$ is an arbitrary $2N$-partite non-signaling distribution then for any ${\bf a} = (a_1, \ldots, a_N)$ we have
\begin{equation}\label{l01}
  P_{{\bf A}|{\bf X}} ({\bf a}, {\bf 0}) = 
	\left(\bigotimes_{n=1}^N \beta_{a_n} \right) \cdot P_{{\bf A}, {\bf B}|{\bf X}, {\bf Y}}\ ,
\end{equation}
where ${\bf 0} =(0,\ldots ,0)$.
\end{lemma}

{\em Proof:} Let us first consider the bound (\ref{l01}) for one pair of systems ($N=1$). The non-signaling constraint $P_{A|X,Y}(0,0,0) = P_{A|X,Y}(0,0,M-1)$ can also be expressed as the scalar product
\begin{equation}\nonumber
  \begin{array}{|cc|cc|cc|cc|}
    \hline
    -1 &-1 &  &  & & & 1 & 1 \\
     0 & 0 &  &  & & & 0 & 0 \\
    \hline
    & & & & & & & \\
    & & & & & & & \\
    \hline
    & &\ddots &\phantom{m} &\ddots &\phantom{m} & &\phantom{-1} \\
    & & & & & & & \\
    \hline
    & & & & & & & \\
    & & & & & & & \\
    \hline
  \end{array} \cdot P_{A,B|X,Y} =0
\end{equation}
and the non-signaling constraint $P_{B|X,Y}(0,0,0) = P_{B|X,Y}(0,1,0)$ can be expressed as
\begin{equation}\nonumber
  \begin{array}{|cc|cc|cc|cc|}
    \hline
    -1 &0 & &\phantom{-1} & &\phantom{-1} &\phantom{\ddots} &\phantom{-1} \\
    -1 &0 & & & & & & \\
    \hline
    1 &0 & & & & & & \\
    1 &0 & & & & & & \\
    \hline
    & &  &  & \ddots & & & \\
    & &  &  & & & & \\
    \hline
    & \phantom{\ddots}&\ddots & &\ddots & & & \\
    & & & & & & & \\
    \hline
    & & & & & & & \\
    & & & & & & & \\
    \hline
  \end{array} \cdot P_{A,B|X,Y} =0\ .
\end{equation}
The remaining non-signaling constraints can be written in an analogous fashion. A linear combination of those equalities gives
\begin{equation}\label{lcc}
  \begin{array}{|cc|cc|cc|cc|}
    \hline
    1 &1 & & & & &\tau &\tau \\
    1 &1 & & & & &\tau &\tau \\
    \hline
    1 &1 &1 &1 & & & & \\
    1 &1 &1 &1 & & &\phantom{\tau_-}&\phantom{\tau_-} \\
    \hline
    \phantom{\ddots} &\phantom{1}  &\ddots & &\ddots & & & \\
      &  & &  &   &   & & \\
    \hline
     & & & &1 &1 &1 &1 \\
     & & & &1 &1 &1 &1 \\
    \hline
  \end{array}
  \cdot P_{A,B|X,Y} =0\ ,
\end{equation}
where $\tau = 1-2M$. If $P_{A,B|X,Y}$ is a non-signaling distribution, the following equalities hold.
\begin{eqnarray*}
  P_{A|X} (0,0) =
  \begin{array}{|cc|cc|cc|cc|}
    \hline
    1 & 1 &\phantom{\ddots} &\phantom{1} &\phantom{\ddots} &\phantom{1} &\phantom{\ddots} &\phantom{1} \\
    0 & 0 & & & & & & \\
    \hline
    & & & & & & & \\
    \phantom{\ddots}& & & & & & & \\
    \hline
    & &\ddots & &\ddots & & & \\
    & & & & & & & \\
    \hline
    & & & & & & & \\
    & & & & & & & \\
   \hline
    \end{array} \cdot P_{A,B|X,Y}
  \\ = \frac{1}{2}
  \begin{array}{|cc|cc|cc|cc|}
    \hline
     0& 1 & & & & & 1 & 1 \\
    -1& 0 & & & & & 0 & 0 \\
    \hline
    1 & 0 & & & & & & \\
    1 & 0 & & & & & & \\
    \hline
     &  & & & \ddots & & & \\
     &  & & & & & & \\
    \hline
     & &\ddots &\phantom{1} &\ddots &\phantom{1} &\phantom{-1} & \\
    & & & & & & & \\
    \hline
    & & & & & & & \\
    & & & & & & & \\
   \hline
    \end{array} \cdot P_{A,B|X,Y}
  \\ = \frac{1}{2}
  \begin{array}{|cc|cc|cc|cc|}
    \hline
     0& 1&  & &\phantom{-1} &\phantom{-1}  & 2  & 1 \\
    -1& 0&  &  &  &  & 1& 0 \\
    \hline
     0&-1& 0& 1&  &  &  & \\
     1& 0&-1& 0&  &  &  & \\
    \hline
      &  &\ddots&  &\ddots  &  &  &  \\
      &  &  &  &  &  &\phantom{\ddots}  & \\
    \hline
      &  &  &  & 0&-1& 0 & 1 \\
      &  &  &  & 1& 0& -1& 0 \\
   \hline
    \end{array} \cdot P_{A,B|X,Y}
\end{eqnarray*}
The second and third equalities follow by adding linear combinations of non-signaling constraints. The above plus (\ref{lcc}) times $1/4M$ gives
\[
  P_{A|X} (0,0) = (\mu +\nu) \cdot P_{A,B|X,Y}\ .
\]
Under the relabeling 
\[
	(A,B)\rightarrow (A\oplus 1,B\oplus 1)\ ,
\]
we have the transformations 
\begin{eqnarray*}
  P_{A|X} (0,0) &\rightarrow& P_{A|X} (1,0)\ , \\
	\mu &\rightarrow& \mu\ , \\
	\nu &\rightarrow& -\nu\ ,
\end{eqnarray*} 
which imply $P_{A|X} (a,0) = \beta_a \cdot P_{A,B|X,Y}$. The generalization to $N$ pairs of systems is straightforward. Each non-signaling constraint involves a linear combination of the entries of $P_{{\bf A}, {\bf B}|{\bf X}, {\bf Y}}$ where all indexes remain constant except the ones corresponding to one system (like for instance $a_1,x_1$). Hence we can apply the above argument to each of the $N$ pairs separately, obtaining (\ref{l01}). \hfill $\Box$

\subsection{Privacy amplification}

The following analysis of privacy amplification is similar to the one in \cite{lluis}, but has the advantage that it is valid for  any choice of error correction scheme. On the other hand, it has the disadvantage that it needs a random hash function $G$, in particular a two-universal one \cite{BBCM94}, while the one in \cite{lluis} works with a deterministic hash function.

%It has the advantage that one can hash out any information about the raw key $C=f({\bf A})$, that is, the function $f$ is arbitrary. Contrary, the scheme introduced in \cite{lluis} only works when the function $f$ is generic. 

\begin{defin}\label{2universal}
	A random function $G:\zu^{N} \rightarrow \zu^{N\s}$ is called two-universal \cite{BBCM94} if for any pair ${\bf a, a'} \in \zu^N$ such that ${\bf a \neq a'}$ we have
\begin{equation}\label{bbb}
	\mathrm{prob} \{G({\bf a}) = G({\bf a}')\} \leq 2^{-N\s}\ .
\end{equation}
\end{defin}

\begin{lemma}\label{gros}
	If $G:\zu^N \rightarrow \zu^{N\s}$ is a two-universal random function, then for any subset ${\cal A} \subseteq \zu^N$ we have 
\begin{equation}\label{gros f}
	\sum_{k,g} P_G (g) \Big| 	\sum_{{\bf a}\in {\cal A}}
	\big( \delta^k_{g({\bf a})} -2^{-N\s}\big) \Big|
 \leq
	\sqrt{2^{N\s} |{\cal A}|}\ ,
\end{equation}
where $k$ runs over $\zu^{N\s}$.
\end{lemma}

\noindent {\em Proof } 
In what follows we take the square of the left-hand side of (\ref{gros f}); use the convexity of the square function; sum over $k$; partially sum over ${\bf a, a'},g$; use the two-universality of $G$; and a trivial bound.
\begin{eqnarray}
\nonumber &&
	\Big( \sum_{k,g} P_G (g) \Big| 	\sum_{{\bf a}\in {\cal A}}
	\big( \delta^k_{g({\bf a})} -2^{-N\s}\big) \Big| \Big)^2
\\ \nonumber &\leq&
	\sum_{k,g} 2^{-N\s} P_G(g) \!\!\sum_{{\bf a,a'} \in {\cal A}}\!\!
	\Big( 2^{2 N\s} \delta^k_{g({\bf a})} \delta^k_{g({\bf a}')}
	+1 -2^{1+N\s} \delta^k_{g({\bf a})} \Big)
\\ \nonumber &=&
	\sum_{g} P_G(g) \!\sum_{{\bf a,a'} \in {\cal A}}\!
	\Big( 2^{N\s}\, \delta^{g({\bf a})}_{g({\bf a}')} -1 \Big)
\\ \nonumber &=&
	2^{N\s} \hspace{-5mm} \sum_{{\bf a,a'} \in {\cal A}:\, {\bf a\neq a'}}\! \Big( \sum_{g} P_G(g)\, 
	\delta^{g({\bf a})}_{g({\bf a}')}\Big) + 2^{N\s}|{\cal A}|
	- |{\cal A}|^2
\\ \nonumber &\leq&
	\big( |{\cal A}|^2 -|{\cal A}| \big) + 2^{N\s}|{\cal A}|
	- |{\cal A}|^2
\\ \nonumber &\leq&
	2^{N\s}|{\cal A}|\ .
\end{eqnarray}
\hfill $\Box$\medskip

\begin{theorem}\label{lluis c2}
Let $P_{{\bf A}, {\bf B}, E| {\bf X}, {\bf Y}, Z}$ be a $(2N\k +1)$-partite non-signaling distribution. Suppose that Alice's systems are measured with ${\bf X}={\bf 0}$, obtaining the outcomes ${\bf A}$. Let $C= f({\bf A})$ where $f:\zu^{N\k} \rightarrow \zu^{N\c}$ is a given function, and $K= G({\bf A})$ where $G:\zu^{N\k} \rightarrow \zu^{N\s}$ is a two-universal random function. Then
%\begin{eqnarray}\nonumber && \sum_{k,c,g} \max_z \sum_e \Big| P_{K,C,E,G|{\bf X}, Z}(k,c,e,g,{\bf 0}, z) -\\ \nonumber && -2^{-N\s} P_{C,E,G|{\bf X}, Z}(c,e,g,{\bf 0}, z) \Big| \\ \label{ll} &\leq& \sqrt{2}^{N+N\s+N\c+1}\, \left\langle \mbox{$\prod_{n=1}^N \left( \frac 1 2 + M W_n\right)$} \right\rangle\ ,\end{eqnarray}
\begin{eqnarray}
	\nonumber
	&& \sum_{k,c,g} \max_z \sum_e 
	\Big| P_{k,c,e,g|z} 
	-2^{-N\s} P_{c,e,g|z} \Big|
	\\ \label{ll}
	&\leq& \sqrt{2}^{N\k+N\s+N\c+1}\, 
	\left\langle \mbox{$\prod_{n=1}^{N\k} \left( \frac 1 2 + M W_n\right)$} \right\rangle\ ,
\end{eqnarray}
where $W_n = (A_n\oplus B_n \oplus \delta_{X_n}^{0} \delta_{Y_n}^{M-1})$ and the expectation in~\eqref{ll} is taken with respect to the distribution $P_{{\bf a}, {\bf b}| {\bf x}, {\bf y}} \prod_{n=1}^{N\k} Q_{x_n ,x_n}$, where $Q_{x_n , y_n}$ is defined in~\eqref{Q}.
\end{theorem}

\noindent {\em Proof } For any subset ${\cal A} \subseteq \zu^{N\k}$ we have the following chain of component-wise inequalities.
\begin{eqnarray}
\nonumber &&
	\sum_{k,g} P_G (g) \Big| 	\sum_{{\bf a}\in {\cal A}}
	\big( \delta^k_{g({\bf a})} -2^{-N\s}\big) 
	\bigotimes_{n=1}^{N\k} \beta_{a_n} \Big|
\\ \nonumber &\preceq&
	\sum_{k,g} P_G (g) \Big(	   
	\mu^{\otimes N\k} \Big| \sum_{{\bf a}\in {\cal A}} 
		\big( \delta^k_{g({\bf a})} -2^{-N\s} \big) \Big| + 
\\ \nonumber && +\
	|\nu|\otimes\mu^{\otimes N\k-1} \Big| \sum_{{\bf a}\in {\cal A}} 
	(-1)^{a_1} \big( \delta^k_{g({\bf a})} -2^{-N\s} \big) \Big| +
\\ \nonumber && +\cdots +
	|\nu|^{\otimes N\k} \Big| \sum_{{\bf a}\in {\cal A}}	
	(-1)^{a_1+\cdots +a_{N\k}}	\big( \delta^k_{g({\bf a})} -2^{-N\s} 
	\big) \Big| \Big)
\\ \nonumber &\preceq& 
	\mu^{\otimes N\k} \sqrt{2^{N\s} |{\cal A}|} + 
	|\nu| \otimes\mu^{\otimes N\k-1} \sqrt{2^{1+ N\s} |{\cal A}|} 
\\ \nonumber &&
	+\cdots +|\nu|^{\otimes N\k} \sqrt{2^{1+N\s}  |{\cal A}|}
\\ \label{ffinal} &\preceq& 
	\sqrt{2^{1+N\s} |{\cal A}|}\ \beta^{\otimes N\k}  \ .
\end{eqnarray}
In the first step we used the expansion
\begin{eqnarray}\label{bigotimes}
	&& \bigotimes_{n=1}^N \beta_{a_n} 
\\ \nonumber &=& 
	\mu^{\otimes N\k} + (-1)^{a_1} \nu\otimes\mu^{\otimes N\k} 
	+\cdots +	(-1)^{a_1+\cdots +a_{N\k}} \nu^{\otimes N\k}\ ,
\end{eqnarray}
as well as  the component-wise triangular inequality. In the second step we used the following triangular inequality for any ${\bf u} \in \zu^{N\k}$
\begin{eqnarray*}
&&
	\Big|	\sum_{{\bf a}\in {\cal A}} (-1)^{{\bf a}\cdot {\bf u}}
	\big( \delta^k_{g({\bf a})} -2^{-N\s}\big) \Big|
\\ \label{gros f2} &\leq&
	\Big| \sum_{{\bf a}\in {\cal A}:\, {\bf a}\cdot {\bf u}=0 \bmod 2} 
	\hspace{-2mm} \big( \delta^k_{g({\bf a})} -2^{-N\s}\big) \Big|
\\ && +\ 
	\Big|	\sum_{{\bf a}\in {\cal A}:\, {\bf a}\cdot {\bf u}=1 \bmod 2} 
	\hspace{-2mm} \big( \delta^k_{g({\bf a})} -2^{-N\s}\big) \Big|\ ,
\end{eqnarray*}
Lemma \ref{gros}, and the concavity of the square-root function
\begin{eqnarray}\label{concavity}
	\sum_{i=1}^{M} \sqrt{t_i} \leq\ \sqrt{ M \sum_{i=1}^{M} t_i }\ .
\end{eqnarray}
For the last inequality all terms are summed up by using $\beta= \mu +|\nu|$.

In the rest of this proof the following notation is used. We denote by $P_{{\bf A,B},e| {\bf X,Y},z} = P_{{\bf A,B} ,E| {\bf X,Y},Z}(e,z)$ the vector with entries $P_{{\bf A,B} ,E| {\bf X,Y},Z}({\bf a,b} ,e, {\bf x,y},z)$ for all values of ${\bf a,b,x,y}$ and fixed values of $e,z$. Following this notation we can write $P_{\bf a} = P_{\bf A} ({\bf a})$. For any subset ${\cal A} \subseteq \zu^{N\k}$ and any set of coefficients $\eta_{\bf a}$ we have the following chain of equalities and inequalities,
\begin{eqnarray}
\nonumber &&
	\sum_e P_{e|z} \Big| 
	\sum_{{\bf a}\in {\cal A}} %\hspace{-2mm} 
	\eta_{\bf a}\, P_{{\bf a}|e,z} \Big|
\\ \nonumber &=&
	\sum_e P_{e|z} \Big|
	\sum_{{\bf a}\in {\cal A}} %\hspace{-2mm} 
	\eta_{\bf a}\,
	\bigotimes_{n=1}^{N\k} \beta_{a_n} \!\cdot P_{{\bf A,B|X,Y},e,z} \Big|
\\ \nonumber &\leq&
	\sum_e P_{e|z} \Big|  
	\sum_{{\bf a}\in {\cal A}} %\hspace{-2mm} 
	\eta_{\bf a}\,
	\bigotimes_{n=1}^{N\k} \beta_{a_n} \Big| \cdot P_{{\bf A,B|X,Y} ,e,z}
\\ \nonumber &=&
	\Big| \sum_{{\bf a}\in {\cal A}} %\hspace{-2mm} 
	\eta_{\bf a}\,
	\bigotimes_{n=1}^{N\k} \beta_{a_n} \Big| \cdot
	\sum_e P_{e|z}\, P_{{\bf A,B|X,Y},e,z}
\\ \label{mezzanine} &=&
	\Big| \sum_{{\bf a}\in {\cal A}} %\hspace{-2mm} 
	\eta_{\bf a}\,
	\bigotimes_{n=1}^{N\k} \beta_{a_n} \Big| \cdot
	P_{\bf A,B|X,Y}\ ,
\end{eqnarray}
where we have respectively used: Lemma \ref{lemma 1}, the convexity of the absolute value function, the linearity of the scalar product, and the definition of the conditional distribution. The following establishes (\ref{ll}).
\begin{eqnarray}
\nonumber	&&
	\sum_{k,c,g} \max_z \sum_e \Big| P_{k,c,g,e|z} - 
	2^{-N\s} P_{c,g,e|z} \Big|
\\ \nonumber &=&
	\sum_{k,c,g} \max_z \sum_e P_{g,e|z }\Big| P_{k,c|g,e,z} - 
	2^{-N\s} P_{c|e,z} \Big|
\\ \nonumber &=&
	\sum_{k,c,g} P_g \max_z \sum_e P_{e|z} \Big| 
	\sum_{{\bf a}\in f^{-1}(c)} \hspace{-2mm} 
	\big( \delta^k_{g({\bf a})} -2^{-N\s} \big) P_{{\bf a}|e,z} \Big|
\\ \nonumber &\leq&
	\sum_{k,c,g} P_{g} \Big|  
	\sum_{{\bf a}\in f^{-1}(c)} \hspace{-2mm} 
	\big( \delta^k_{g ({\bf a})} -2^{-N\s} \big) 
	\bigotimes_{n=1}^{N\k} \beta_{a_n} \Big| \cdot
	P_{\bf A,B|X,Y}
\\ \nonumber &\leq&
	\sum_{c} \sqrt{2^{1+N\s} |f^{-1}(c)|}\ 
	\beta^{\otimes N\k}\hspace{-1mm} \cdot	P_{\bf A,B|X,Y}
\\ \label{muji} &\leq&
	\sqrt{2^{1+N\s +N\c +N\k}}\, \beta^{\otimes N\k}\cdot 
	P_{{\bf A},{\bf B}|{\bf X},{\bf Y}}\ ,
\end{eqnarray}
In the above we have respectively used: the definition of conditional distribution and the fact that $C,E,Z$ are independent from $G$; equality $P_{c} = \sum_{{\bf a}\in f^{-1}(c) } P_{\bf a}$ and the independence of ${\bf A},E,Z$ from $G$; inequality (\ref{mezzanine}) with ${\cal A} = f^{-1} (c)$; the component-wise inequality (\ref{ffinal}) together with the fact that the components of the vector $P_{\bf A,B|X,Y}$ are positive; and the last inequality follows from (\ref{concavity}) and $\sum_c |f^{-1}(c)| = 2^{N\k}$.
\hfill $\Box$

\subsection{Security from estimated information}

According to the previous theorem, the security of the secret key can be bounded in terms of the quantity 
\begin{equation}\label{e59}
  \beta^{\otimes N\k}\cdot P_{{\bf A}\k,{\bf B}\k|{\bf X}\k,{\bf Y}\k}
  = \left\langle 
  \prod_{n=1}^{N\k}\! \left( \frac 1 2 + M W_n\right)
  \right\rangle,
\end{equation}
which does not depend on $E$ at all, but only on the distribution of data held by the honest parties! This is a particular manifestation of the monogamy of non-local correlations. However, also the distribution $P_{{\bf A}\k, {\bf B}\k | {\bf X}\k, {\bf Y}\k}$ that occurs in~\eqref{e59}  is not necessarily known. Hence, in order to be of use, we need to relate it to an observable quantity, such as $\bar W$, defined in (\ref{est non-locality}). This is the purpose of Lemma~\ref{uc_PA} below. 

To simplify notation, we restrict to the relevant pairs of systems, that is, the ones that are used to either estimate the amount of non-locality $n\in {\cal N}\e$, or the ones constituting the raw key $n\in {\cal N}\k$. These are the ones that are not discarded in the protocol.

%\RR{I was wondering whether there is a way to present the technical claims in a more modular way, separating the estimation from the actual privacy amplification, as it is usually doen in ``conventional'' QKD. But I guess there is a reason why this is all combined in one lemma, isn't there?}

\begin{lemma}\label{uc_PA}
Let $N\k$ and $N\e$ be two positive integers and $P_{{\bf A}, {\bf B}, E| {\bf X}, {\bf Y},Z}$ a $(2 N_{\rm u} +1)$-partite non-signaling distribution, where $N_{\rm u}= N\k +N\e$. Let the random variable ${\bf H}= (H_1, \ldots, H_{N_{\rm u}})$ be independent from ${\bf A,B,X,Y},E,Z$, and uniformly distributed over the strings $\zu^{N_{\rm u}}$ with $N\k$ zeroes and $N\e$ ones. Suppose that all Alice's systems $n\in \{1, \ldots, N\}$ with $H_n=0$ are measured with $X_n=0$ obtaining the $N\k$-bit outcome ${\bf A}\k$. Suppose the $N\e$ pairs with $H_n=1$ are measured with $(X_n,Y_n)$ following the distribution $Q_{X,Y}$ defined in~\eqref{Q}, obtaining the outcome ${\bf U}= [(A_n, B_n, X_n, Y_n) : H_n=1]$. This is used to compute the variable
\begin{equation}
  \bar W = \frac{1}{N\e} \sum_{n: H_n=1}
  (A_n\oplus B_n \oplus \delta_{X_n}^{0} \delta_{Y_n}^{M-1})	\ .
\end{equation}
Let $C= f({\bf A}\k)$ where $f:\zu^{N\k} \rightarrow \zu^{N\c}$ is a given function. If $G:\zu^{N\k} \rightarrow \zu^{N\s}$ is a two-universal random function with output size
\begin{eqnarray}\nonumber
  N\s (\bar w) &=& \max_{\theta \in [0,1]} \big[ 2 N\e D(\bar w\|\theta) -2 N\k \log_2 (1/2 + M\theta)\big]
\\ \label{assign Ns uc} &&
   -N\k -N\c  -2 \log_2 (8 N\e N_{\rm u} /\epsilon) \ ,
\end{eqnarray}
and $K= G({\bf A}\k)$ then
%\begin{eqnarray}\label{theorem 18}\nonumber && \sum_{k,\bar w,c,g} \max_z \sum_e \Big| P_{k,\bar w,c,e,g|z} \\ \label{cosss} && -\, 2^{-N\s (\bar w)} P_{\bar w,c,e,g|z} \Big| \ \leq\ \epsilon\ . \end{eqnarray}
\begin{equation}\label{theorem 18}
  \sum_{k,{\bf h}, {\bf u},c,g} \max_z \sum_e \Big| P_{k,{\bf h}, {\bf u},c,e,g|z} -2^{-N\s (\bar w)} P_{{\bf h}, {\bf u},c,e,g|z} \Big|
	\ \leq\ \epsilon
\end{equation}
holds for any $\epsilon >0$.
\end{lemma}

\noindent {\em Proof } 	
For each value of ${\bf h}$ define the disjoint sets 
\begin{eqnarray*}
  {\cal N}\k^{\bf h} &=& \{n: h_n=0\}\ ,
\\
  {\cal N}\e^{\bf h} &=& \{n: h_n=1\}\ ,
\end{eqnarray*}
satisfying ${\cal N}\k^{\bf h} \cup {\cal N}\e^{\bf h} = \{1, \ldots, N_{\rm u} \}$. Note that these are the same sets as~\eqref{condition} and~\eqref{raw set}. We also define
\begin{eqnarray*}
  W_n &=& [A_n\oplus B_n \oplus \delta_{X_n}^{0}
  \delta_{Y_n}^{M-1}] \mbox{ for } n=1,\ldots ,N_{\rm u},
\\
  {\bf W} &=& (W_{1}, \ldots, W_{N_{\rm u}}) \ ,
\\
  {\bf W}\k &=& (W_n : n\in {\cal N}\k^{\bf h})\ ,
\\
  {\bf W}\e &=& (W_n : n\in {\cal N}\e^{\bf h})\ ,
\\
  {\bf U} &=& \left[ (A_n, B_n, X_n, Y_n) : 
  n\in {\cal N}\e^{\bf h} \right]\ .
\end{eqnarray*}
The distribution for ${\bf W}$ is
\[
  P_{\bf w}= \sum_{\bf a,b,x,y} \!\!
  P_{\bf a,b|x,y} 
  \prod_{n=1}^{N_{\rm u}} Q_{x_n,y_n}\, 
  \delta_{[a_n\oplus b_n \oplus \delta_{x_n}^{0} \delta_{y_n}^{M-1}]}^{w_n}\ ,
\]
where $P_{\bf A,B|X,Y}$ is the marginal of $P_{{\bf A}, {\bf B}, E| {\bf X}, {\bf Y},Z}$ and $Q_{X,Y}$ is defined in~\eqref{Q}. 
%For a fixed ${\bf h}$, the joint distribution for ${\bf W}\k$ and $\bar W$ is \begin{equation}\label{Pbar} P_{{\bf w}\k, \bar w}= \sum_{\bf w\e} P_{\bf w}\,  \delta_{[\sum_{n\in {\cal N}\e^{\bf h}} w_n /N\e]}^{\bar w}\ .\end{equation}
For what follows it is also useful to define the symmetrized version of $P_{\bf w}$, namely
\begin{equation}\label{symp}
  P^{\rm sym}_{w_1, \ldots, w_{N_{\rm u}}} = 
  \sum_{\pi} \frac{1}{N_{\rm u} !}
  P_{w_{\pi(1)}, \ldots, w_{\pi(N_{\rm u})}}\ ,
\end{equation}
where $\pi$ runs over the permutations of the symbols $\{1, \ldots, N_{\rm u}\}$.

For each value of ${\bf h}$ and ${\bf u}$ the conditioned distribution $P_{{\bf A}\k, {\bf B}\k, E| {\bf X}\k, {\bf Y}\k, Z, {\bf h}, {\bf u}}$ is $(2N\k +1)$-partite non-signaling, hence, Theorem~\ref{lluis c2} applies to it. By definition, the random variable ${\bf H}$ is independent from $Z$; and by no-signaling, the random variable ${\bf U}$ is independent from $Z$. Hence, $P_{{\bf h}, {\bf u}|z} = P_{{\bf h}, {\bf u}}$.
This allows for taking the common factor $P_{{\bf h}, {\bf u}}$ out of the absolute value in~\eqref{theorem 18}, and applying Theorem~\ref{lluis c2} to each term, obtaining:
\begin{eqnarray}
\nonumber &&
  \sum_{k,{\bf h}, {\bf u},c,g} \max_z \sum_e 
  \Big| P_{k,{\bf h}, {\bf u},c,e,g|z} 
  -2^{-N\s (\bar w)} P_{{\bf h}, {\bf u},c,e,g|z} \Big|
\\ \nonumber &=&
  \sum_{k,{\bf h}, {\bf u},c,g} \max_z P_{{\bf h}, {\bf u}|z} \sum_e
  \Big| P_{k,c,e,g|z} 
  -2^{-N\s (\bar w)} P_{c,e,g|z} \Big|
\\ \nonumber &=&
  \sum_{{\bf h}, {\bf u}} 
  P_{{\bf h}, {\bf u}} 
\\ \nonumber &&
  \times \sum_{k,c,g} \max_z \sum_e 
  \Big| P_{k,c,e,g|z, {\bf h}, {\bf u}} - 
  2^{-N\s (\bar w)} P_{c,e,g|z, {\bf h}, {\bf u}} \Big|
\\ \nonumber &\leq &
  \sum_{{\bf h}, {\bf u}} P_{{\bf h}, {\bf u}}\, \sqrt{2}^{N\k + N\s (\bar w) + N\c +1}
\\ \nonumber && \times
  	\sum_{\bf w\k} P_{{\bf w}\k|{\bf h}, {\bf u}}   	\!\!\prod_{n\in {\cal N}\k^{\bf h}}\!\! \left( \frac 1 2 + M w_n\right)
\\ \label{no65} &= &
  \sum_{\bf h, w} P_{{\bf h}, \bf w}\, \sqrt{2}^{N\k + N\s (\bar w) + N\c +1}
  \!\! \prod_{n\in {\cal N}\k^{\bf h}}\!\! \left( \frac 1 2 + M w_n\right)
\end{eqnarray}
The last equality follows from the fact that $\bar w$ is a function of ${\bf w}\e$ which in turn is a function of ${\bf u}$. Hence, averaging over $({\bf w}\k, {\bf u})$ is equivalent to averaging over $({\bf w}\k, {\bf w}\e)= {\bf w}$.

Now, let $\pi\k$ be any permutation of the variables $w_n$ with $n\in {\cal N}\k^{\bf h}$, and $\pi\e$ any permutation of the variables $w_n$ with $n\in {\cal N}\e^{\bf h}$.
The fact that $\bar w$ and $\prod_{n\in {\cal N}\k^{\bf h}}\!\! \left( \frac 1 2 + M w_n\right)$ are invariant under any permutation $\pi\k$ and $\pi\e$ implies that~\eqref{no65} is equal to
\begin{eqnarray}
\nonumber &&
  \sum_{{\bf h, w}, \pi\k, \pi\e } 
  \frac{P_{{\bf h}, (\pi\k \pi\e{\bf w})}}
  {N\k ! N\e !}
  \, \sqrt{2}^{N\k + N\s (\bar w) + N\c +1}
\\ \nonumber && \times\!\!
  \prod_{n\in {\cal N}\k^{\bf h}}\!\! \left( \frac 1 2 + M w_n\right)
\\ \label{pw} &= &
  \sum_{\bf w} P^{\rm sym}_{\bf w}\, \sqrt{2}^{N\k + N\s (\bar w) + N\c +1}
  \!\! \prod_{n\in {\cal N}\k^{\bf h}}\!\! \left( \frac 1 2 + M w_n\right)\ \ \ \ \ \ 
\end{eqnarray}
The above equality follows from noting that the average over all permutations $\pi\k, \pi\e$ combined with the average over all partitions $({\cal N}\k^{\bf h} ,{\cal N}\e^{\bf h})$ of $\{1, \ldots, N_{\rm u}\}$ (or equivalently the average over ${\bf h}$) is equivalent to the average in~\eqref{symp}.

Since $P^{\rm sym}_{\bf w}$ is symmetric we can apply Lemma~\ref{prop 1} to upper-bound~\eqref{pw} in terms of a maximization over i.i.d. distributions $P_{w_1, \ldots, w_{N_\mathrm{u}}}^{\rm iid} = P_{w_1} \cdots P_{w_{N_\mathrm{u}}}$. These i.i.d. distributions are parametrized by the single number $\theta= P_W (1) \in [0,1]$. The following is an upper bound for~\eqref{pw}.
\begin{eqnarray*}
&&
  (N_\mathrm{u} +1) \,
  \max_{P_{\bf W}^{\rm iid}}\,
  \sum_{\bf w} P_{\bf w}^{\rm iid}\, 
  \sqrt{2}^{N\k + N\s (\bar w) + N\c +1}
\\ && \times
  \prod_{n\in {\cal N}\k ^{\bf h}} \left( \frac 1 2 + M w_n\right)
\\ &=&
  (N_\mathrm{u} +1) \, \max_{\theta \in [0,1]}\, 
  \sum_{\bar w} P_{\bar w}^{(\theta)}
\\ &&\times
  \sqrt{2}^{N\k + N\s (\bar w) + N\c +1}
  \left( \frac 1 2 + M \theta\right)^{N\k}
\end{eqnarray*}
To obtain the above equality we use
\begin{equation}\nonumber
  \sum_{\bf w\k} P^{\rm iid}_{\bf w\k}\,
  \prod_{n=1}^{N\k} \left( \frac 1 2 + M w_n\right)
  =
  \left( \frac 1 2 + M \theta\right)^{N\k} ,
\end{equation}
and express the average over ${\bf w\e}$ in terms of the distribution $P_{\bar w}^{(\theta)}$, which is 
\begin{equation}
  P_{\bar w}^{(\theta)} = 
  \left(\begin{array}{c} 
    N\e \\ N\e \bar w\end{array}\right)
  \theta^{N\e \bar w} (1-\theta)^{N\e (1-\bar w)}\ .
\end{equation}
It is well known~\cite{TC} that the above distribution can be bounded as
\begin{equation}
  P_{\bar w}^{(\theta)} \leq
  2^{-N\e D(\bar w \| \theta)}\ ,
\end{equation}
where $D(\bar w \| \theta)$ is the binary relative entropy defined in~\eqref{relent}. Putting everything together we obtain
\begin{eqnarray}\nonumber
  &&\sum_{k,{\bf h}, {\bf u},c,g} \max_z \sum_e \Big| P_{k,{\bf h},{\bf u},c,e,g|z} -2^{-N\s (\bar w)} P_{{\bf h},{\bf u},c,e,g|z} \Big|
\\ \nonumber &\leq &
  (N_\mathrm{u} +1)\, \max_{\theta \in [0,1]}\,
  \sum_{\bar w} 2^{(N\k + N\s (\bar w) + N\c +1)/2}
\\ \nonumber && \times
  2^{-N\e D(\bar w\| \theta) +N\k \log (1/2 + M \theta)}
  \vspace{2mm}
\\ \nonumber &\leq &
  (N_\mathrm{u} +1) \max_{\theta \in [0,1]}\,
  \sum_{\bar w} 2^{1/2 -\log_2(8 N\e N_\mathrm{u}/\epsilon)} 
\\  &= &
  (N_\mathrm{u} +1) (N\e+1)\, \frac{\sqrt{2}\, \epsilon}{8 N\e N_\mathrm{u}} 
  \ \leq\ \epsilon\ ,
\end{eqnarray}
where we have used $\sum_{\bar w} 1= N\e+1$.
\hfill $\Box$\medskip

To avoid confusion in the following Theorem, we recall that the the alphabets of the random variables ${\bf A}\k, {\bf B}'\k$ and $G$ depend on the value of $T$, defined in~\eqref{T} or~\eqref{T2}. Particularly, ${\bf A}\k, {\bf B}'\k$ take values in $\zu^{N\k}$, and $G$ takes values in the set of functions $\zu^{N\k} \to \zu^{N\s}$. But the number $N\k$ is a function of ${\bf I,J}$, namely, the number of times $I_n=J_n=0$. And as can be seen in~\eqref{Ns 12}, the quantity $N\s (\bar W)$ is a function of $\bar W$, which in turn is a function of $T$. 

\begin{theorem}
At the end of the protocol described in Section~\ref{protoc}, Alice holds $K_A$, Bob holds $K_B$, and the adversary has all the information
\begin{equation}\label{T2}
  T= \Big[ {\bf I,J}, C,G, (A_n, B_n, X_n, Y_n)\ \forall n\in {\cal N}\e \Big]
\end{equation}
and the system associated to $E,Z$. If the error correction scheme has error probability
\begin{equation}
  \sum_{t, \bf a\k \neq b\k'} 
  P_{t, \bf a\k, b'\k}  \ \leq\  
  \epsilon_\mathrm{erco}
\end{equation}
then we have
\begin{eqnarray}
\nonumber &&
  \hspace{-3mm} \sum_{k_A, k_B, t} 
  \hspace{-1mm} \max_z \sum_e 
  \Big| P_{k_A, k_B,t,e|z} -2^{- N\s (\bar w)}\, 
  \delta_{k_A}^{k_B}\,  
  P_{t,e|z} \Big|
\\ \label{theorem 9} &\leq &
	\epsilon + 2\epsilon_\mathrm{erco}\ . 
\end{eqnarray}
\end{theorem}

\noindent {\em Proof } 	
Using the triangular inequality
\begin{eqnarray}
\nonumber &&
  \Big| P_{k_A, k_B, t,e|z} -2^{- N\s (\bar w)}
  \, \delta_{k_A}^{k_B}\,  
  P_{t,e|z} \Big|
\\ \nonumber &\leq&
  \Big| P_{k_A, k_B,t,e|z} - 
  P_{k_A,t,e|z}\, \delta_{k_A}^{k_B} \Big|
\\ \label{E71} && +
  \Big| P_{k_A, t,e|z}\, \delta_{k_A}^{k_B} 
  -2^{- N\s (\bar w)} 
  \, \delta_{k_A}^{k_B}\, 
  P_{t,e|z} \Big|
\end{eqnarray}
we can bound the left-hand side of~\eqref{theorem 9} with the corresponding two terms. The first term can be simplified by splitting the sum into the terms with $k_A=k_B$ and $k_A\neq k_B$, and using the no-signaling constraint $\sum_e P_{k_A, k_B, t,e|z} =P_{k_A, k_B, t}$, that is
\begin{eqnarray}
\nonumber &&
  \sum_{k_A, k_B,t} 
  \max_z \sum_e 
  \Big| P_{k_A, k_B,t,e|z} - 
  P_{k_A, t,e|z}\, \delta_{k_A}^{k_B} \Big|
\\ \nonumber &=&
  \sum_{k_A \neq k_B,t} 
  \max_z \sum_e 
  P_{k_A, k_B, t,e|z}
\\ \nonumber && +
  \sum_{k_A, t} 
  \max_z \sum_e 
  \Big| P_{k_A, k_B=k_A, t,e|z} - 
  P_{k_A,t,e|z} \Big|
\\ \nonumber &\leq & 
  \sum_{k_A \neq k_B,t} 
  \max_z\ 
  P_{k_A, k_B, t}
\\ \nonumber && +
  \sum_{k_A, t} 
  \max_z \sum_e \left( 
  P_{k_A, t,e|z} - P_{k_A, k_B=k_A, t,e|z}   
  \right)
\\ \nonumber &\leq & 
  \epsilon_\mathrm{erco} +
  \sum_{k_A, t} \left( 
  P_{k_A, t} - P_{k_A, k_B=k_A, t}   
  \right)
\\ \label{ff} &\leq & 
  \epsilon_\mathrm{erco} + 1 - 
  (1- \epsilon_\mathrm{erco})
  = 2 \epsilon_\mathrm{erco}\ .
\end{eqnarray}
In the last two inequalities we have used that
\begin{eqnarray}
\nonumber &&
  \sum_{k_A \neq k_B, t} P_{k_A, k_B, t}
\\ \nonumber &=& 
  \sum_{k_A \neq k_B, 
  t, \bf a\k, b'\k} 
  P_{t, \bf a\k, b'\k}\, \delta_{g(\bf a\k)}^{k_A}
  \, \delta_{g( \bf b'\k)}^{k_B}
\\  &\leq & 
  \sum_{t, \bf a\k \neq b'\k} 
  P_{t, \bf a\k, b'\k} \ \leq\ \epsilon_{\rm erco}\ .
\end{eqnarray}
To bound the second term in \eqref{E71} we invoke Lemma~\ref{uc_PA}. However, in Lemma~\ref{uc_PA} the values of $N\k, N\e$ and the set ${\cal N}_{\rm u} = {\cal N}\k \cup {\cal N}\e$ are fixed, while here this is not the case. To overcome this problem, we can separate the sum over $N\k, N\e, {\cal N}_{\rm u}$ and independently apply Lemma~\ref{uc_PA} to each term with a fixed value of $N\k, N\e, {\cal N}_{\rm u}$. 
\begin{eqnarray}
\nonumber &&
  \sum_{k_A, k_B, t} \max_z \sum_e \delta_{k_A}^{k_B}
  \Big| P_{k_A, t,e|z} 
  -2^{- N\s (\bar w)} 
  P_{t,e|z} \Big|
\\ \nonumber &=&
  \sum_{N\k, N\e, {\cal N}_{\rm u}}
  P(N\k, N\e, {\cal N}_{\rm u})
  \!\! \sum_{k,{\bf h}, {\bf u},c,g} 
  \!\! \max_z \sum_e  
\\ \nonumber &&
  \Big| P_{k,{\bf h}, {\bf u},c,e,g|z,N\k, N\e, {\cal N}_{\rm u}} 
  -2^{-N\s (\bar w)} 
  P_{{\bf h}, {\bf u},c,e,g|z,N\k, N\e, {\cal N}_{\rm u}} \Big|
\\ \label{f} &\leq &
  \sum_{N\k, N\e, {\cal N}_{\rm u}}
  P(N\k, N\e, {\cal N}_{\rm u})\ 
  \epsilon \ =\  \epsilon\ .
\end{eqnarray}
To understand the above equality note that, except for the discarded pairs (with $I_n \neq J_n$) which are not considered in Lemma~\ref{uc_PA}, all the information contained in $t$ is also contained in $[N\k, N\e, {\cal N}_{\rm u}, {\bf h, u}, c,g]$. More specifically, the set ${\cal N}_{\rm u}$ tells us which of the pairs satisfy $I_n = J_n$, and among those, ${\bf h}$ designates the ones with $I_n =0$ and $I_n =1$. Additionally, the information $(A_n, B_n, X_n, Y_n)\ \forall n\in {\cal N}\e$ is contained in ${\bf u}$.

Finally, the combination of~\eqref{ff} and~\eqref{f} gives~\eqref{theorem 9}. \hfill $\Box$

\section{Conclusions}\label{conc}

We have showed that it is possible to generate secret key from correlations that violate the Braunstein-Caves inequality \cite{BC} by a sufficient amount. We proved this according to the strongest notion of security, the so-called universally-composable security \cite{RK,BO}. The only assumption used in the security proof is that, when measuring a system, the outcome does not depend on the choice of observables measured on other systems. One clean (although expensive) way to achieve this within our device-independent scenario would be to use a separate isolated measurement device for each of the measurements. A more practical (but not fully device-independent) variant is to use a single device per party, whose design is chosen such that it can reasonably be assumed that subsequent measurements are independent of each other (i.e., the device should not have any internal memory).

On the technical level, we introduced a scheme for estimating symmetric properties of general probability distributions. Applied to our setting, this allows Alice and Bob to treat arbitrary and unknown correlations as if they were independent and identically-distributed samples. This may be more generally useful to quantify Bell-inequality violations without the i.i.d.~assumption.

This work is inspired by, but goes beyond the philosophy of~\cite{E91} in which the validity of quantum mechanics, in particular, Tsirelson's bound~\cite{cirelson}, is still assumed.  In contrast, \emph{all} we assume is no-signaling. This idea, proposed in~\cite{BHK}, is conceptually simpler. Nevertheless, we hope that our results also contribute to the understanding of the more practical scenario of device-independent quantum cryptography where quantum theory is assumed to hold, but where the honest users still do not have complete control of their quantum apparatuses, or distrust them \cite{MayersYao, AGM, ABGMPV}. In this case, our techniques may still be applied. Furthermore, with appropriate modifications (cf.\ Appendix~\ref{mnc}), it is possible to obtain higher key rates, compared to the pure non-signaling scenario.

%QKD is a present-day technology. Entanglement-based protocols are usually implemented with a source that sequentially sends entangled pairs of systems to Alice and Bob. Each pair is measured in Alice and Bob's locations with the same two apparatuses. In the device-independent scenario, those measuring apparatuses could generate outcomes depending on previous inputs. If this is the case, our assumptions for the security proof do not hold, because there could be signaling between measuring events within the same lab. Hence, in order to use our security proof in a setup that sequentially measures pairs of systems with the same apparatuses, one needs to assume that such apparatuses have no internal memory. It would be desirable to have a security proof which does not need the no-memory assumption. Therefore, an important open problem is to obtain a security proof that only relies on the assumption of non-signaling to the past.

\section{Acknowledgements}

LM acknowledges support from CatalunyaCaixa, the EU ERC Advanced Grant NLST (PHYS RQ8784), EU Qessence project and the Templeton Foundation.
RR acknowledges support by the Swiss National Science Foundation (SNF) through the National Centre of Competence in Research ``Quantum Science and Technology (QSIT)''  and through grant No. 200020-135048, the European Research Council (grant 258932), and the CHIST-ERA project DIQIP.
MC acknowledges financial support by the German Science Foundation (grant CH 843/2-1), the Swiss National Science Foundation (grants PP00P2-128455, 20CH21-138799 (CHIST-ERA project CQC)), the Swiss National Center of Competence in Research ``QSIT'', the Swiss State Secretariat for Education and Research supporting COST action MP1006, and the European Research Council under the European Union's Seventh Framework Program (FP/2007-2013) (grant 337603).
AW acknowledges support by the EC STREP “QCS”, the ERC (Advanced Grant “IRQUAT”), and the Philip Leverhulme Trust.
JB is supported by the EPSRC, and the CHIST-ERA DIQIP projects.

%LlM is supported by Caixa Manresa, the spanish MEC projects (FIS2008-00784 ``TOQATA", FIS2007-60182, Consolider Ingenio 2010 ``QOIT"), and the EU-IP programme ``SCALA". RR is supported by the Swiss National Science Foundation, grant No. 200021-119868. MC is supported by the Swiss National Science Foundation (grant PP00P2-128455) and the German Science Foundation (grants CH 843/1-1 and CH 843/2-1). AW is supported by the U.K. EPSRC through the ``QIP IRC" and an Advanced Fellowship, by a Royal Society Wolfson Merit Award, a Philip Leverhulme Prize, by the European Commission through IP ``QAP", and by the Singapore Ministry of Education. JB is supported by an EPSRC Career Acceleration Fellowship.

\appendix

\section{Monogamy of non-local correlations}\label{mnc}

The following lemma is not necessary for the security proof, but we include it for the following two reasons. First, it provides insight on phenomenon of monogamy, that is, the trade-off between Bell-inequality violation and correlation with a third party. Second, it allows to highly improve the secret key rate of our protocol if we additionally assume that the global initial distribution $P_{{\bf A,B},E|{\bf X,Y},Z}$ is compatible with quantum theory~\cite{Pironioetal}. Note that this extra assumption does not invalidate the fact that the security is device independent.

In order to incorporate the validity of quantum theory as an extra assumption, we can use the results in~\cite{min entropy}, which provide a bound for the security of the secret key in terms of the guessing probability of the raw key. And this is the quantity addressed by the following lemma. 

\begin{lemma}\label{lemma 0}
Let $P_{{\bf A}, {\bf B},E|{\bf X}, {\bf Y}, Z}$ be an arbitrary $(2N+1)$-partite non-signaling distribution and define 
\begin{equation}\label{Pmax}
	\mathcal{P}_\mathrm{guess} ({\bf A}|E, {\bf x}) 
= 
	\max_z \sum_e \max_{\bf a} 
	P_{{\bf A},E| {\bf X},Z} ({\bf a},e, {\bf x},z )\ .
\end{equation}
For any ${\bf x}$ we have
\begin{equation}\label{l1}
  \mathcal{P}_\mathrm{guess} ({\bf A}|E, {\bf x}) \leq\ 
  \left\langle \mbox{$\prod_{n=1}^{N} \left( \frac 1 2 + M W_n\right)$} \right\rangle\ ,
\end{equation}
where the expectation of $W_n = (A_n \oplus B_n \oplus \delta_{X_n}^{0} \delta_{Y_n}^{M-1})$ is taken with the distribution $Q_{X_n,Y_n}$ defined in~\eqref{Q}.
\end{lemma}
{\em Proof} First, note that by using the non-signaling condition we can write
\begin{equation}\label{eq:marginal}
  P_{\bf A,B|X,Y} = \sum_e P_{E|Z}(e,z)\, P_{{\bf A},{\bf B}|{\bf X},{\bf Y}, E,Z}(e,z)\ .
\end{equation}
Now, let us show that
\begin{equation}\label{ineq beta}
	\beta_{a_1} \otimes\cdots\otimes \beta_{a_n} \preceq \beta^{\otimes n}\ ,
\end{equation}
for any $n$ and any $(a_1,\ldots , a_n) \in \zu^n$. First, expand each side of this inequality according to definitions \eqref{beta a} and \eqref{beta}; second, note that $\mp \nu^{\otimes n} \preceq |\nu^{\otimes n}| = |\nu|^{\otimes n}$; and finally, use this to show that each term in the left is component-wise bounded by the corresponding term in the right. Let us show \eqref{l1} for the case ${\bf x}= (0,\ldots,0)$. In the following chain of equalities and inequalities we use, respectively: the definition of $\mathcal{P}_\mathrm{guess}$ in (\ref{Pmax}); Lemma \ref{lemma 1}; inequality \eqref{ineq beta} and positivity of the vectors $P_{{\bf A},{\bf B}|{\bf X},{\bf Y},E,Z} (e,z)$; the linearity of the scalar product; decomposition (\ref{eq:marginal}); and the identity \eqref{id BC}.
\begin{eqnarray*}
  && \mathcal{P}_\mathrm{guess} ({\bf A}|E, {\bf x})
	\\
  &=& \max_z \sum_{e} P_{E|Z}(e,z) \max_{\bf a} P_{{\bf A}|{\bf X},E,Z} ({\bf a},{\bf x},e,z)
  \nonumber
  \\ \nonumber
  &=& \max_z \sum_{e} P_{E|Z}(e,z) \max_{\bf a} \left( \bigotimes_{n=1}^N \beta_{a_n} \right)\!\cdot
	P_{{\bf A},{\bf B}|{\bf X},{\bf Y},E,Z} (e,z)
  \\ \nonumber
  &\leq& \max_z \sum_{e} P_{E|Z}(e,z)\, \beta^{\otimes N}\!\cdot
	P_{{\bf A},{\bf B}|{\bf X},{\bf Y},E,Z} (e,z)
  \\ \nonumber
  &=& \max_z \beta^{\otimes N}\!\cdot \left( \sum_{e} P_{E|Z}(e,z) 
	P_{{\bf A},{\bf B}|{\bf X},{\bf Y},E,Z} (e,z) \right)
  \\ \nonumber
  &=& \beta^{\otimes N}\!\cdot P_{\bf A,B|X,Y}
  \\ \nonumber
  &=&   \left\langle \mbox{$\prod_{n=1}^{N} \left( \frac 1 2 + M W_n\right)$} \right\rangle
\end{eqnarray*}
In order to extend this inequality to all values of ${\bf x}$, consider the relabeling. For any $m \in \{0,\ldots, M-1\}$
\begin{equation}\label{relabeling}
  \begin{array}{lll}
    X & \rightarrow & X + m \bmod M \\
    Y & \rightarrow & Y + m \bmod M \\
    A & \rightarrow & A \oplus I\{M-m \leq X \leq M-1\} \\
    B & \rightarrow & B \oplus I\{M-m \leq Y \leq M-1\}
  \end{array}\ .
\end{equation}
This relabeling corresponds to a permutation of the entries of the vectors \eqref{arrangement} such that
\[
	P_{A|X}(a,0) \rightarrow  P_{A|X}(a,m)\ .
\]
This relabeling leaves the vector $\beta$ invariant. Hence, performing the relabeling to each pair with $m=x_n$, the above inequality for ${\bf x}= (0,\ldots,0)$ is generalized to any value of ${\bf x}$.
\hfill $\Box$

\end{document}